\documentclass[aps,twocolumn,showpacs,%
superscriptaddress,unsortedaddress,floatfix]{revtex4}
\usepackage{graphicx}%
\usepackage{amsmath}
\usepackage{bm}

\let\d=\delta%
\let\veps=\varepsilon%
\let\k=\kappa%
\let\w=\omega%
\let\what=\widehat%
\let\n=\nonumber%
\let\tol=\leftarrow%
\newcommand\avg[1]{\left\langle\textstyle#1\right\rangle}%
\newcommand\set[1]{\left\{\textstyle#1\right\}}%
\newcommand\ket[1]{\left|\textstyle#1\right\rangle}%
\newcommand\bra[1]{\left\langle\textstyle#1\right|}%
\newcommand\braket[1]{\left\langle\textstyle#1\right\rangle}%
\newcommand\whatbf[1]{\what{\mathbf{\textstyle#1}}} %
\newcommand\hatbf[1]{\hat{\mathbf{\textstyle#1}}} %
\newcommand{\be}{\begin{equation}}%
\newcommand{\ee}{\end{equation}}%
\newcommand{\bem}{\begin{multline}}%
\newcommand{\eem}{\end{multline}}%
\newcommand{\bea}{\begin{eqnarray}}%
\newcommand{\eea}{\end{eqnarray}}%
\begin{document}
\bibliographystyle{apsrev}

\title{Nonequilibrium plasmons and transport properties of a
double--junction quantum wire}

\author{Jaeuk U. Kim}
\affiliation{ Department of Physics, G\"oteborg University, SE-412
96 Gothenburg, Sweden}
\author{Mahn-Soo Choi}
\affiliation{Department of Physics, Korea University, Seoul
136-701,   Korea}%
\author{Ilya V. Krive}
\affiliation{ Department of Applied Physics, Chalmers University of
Technology, SE-412 96 Gothenburg, Sweden} \affiliation{
   B.I. Verkin Institute for Low Temperature Physics and Engineering,
   Lenin Avenue, 47, Kharkov 61103, Ukraine}
\author{Jari M. Kinaret}
\affiliation{ Department of Applied Physics, Chalmers University of
Technology, SE-412 96 Gothenburg, Sweden}

\date{\today}

\begin{abstract}
We study theoretically the current-voltage characteristics, shot
noise, and full counting statistics of a quantum wire double
barrier structure. We model each wire segment by a spinless
Luttinger liquid. Within the sequential tunneling approach, we
describe the system's dynamics using a master equation. We show
that at finite bias the non-equilibrium distribution of plasmons
in the central wire segment leads to increased average current,
enhanced shot noise, and full counting statistics corresponding to
a super-Poissonian process. These effects are particularly
pronounced in the strong interaction regime, while in the
non-interacting case we recover results obtained earlier using
detailed balance arguments.
\end{abstract}


\maketitle

\section{Introduction}

The recent discovery of novel one-dimensional (1D) conductors that
do not follow Fermi liquid theory has inspired extensive research
activities both in theory and experiment. The generic behavior of
electrons in 1D conductors is well described by the Luttinger
liquid (LL) theory which is a generalization of the
Tomonaga-Luttinger (TL) model \cite{
Tomonaga50a,Luttinger63a,Haldane81b}.
 In one-dimensional conductors, unlike their higher dimensional
counterparts, electron-electron ($e-e$) Coulomb interaction is
poorly screened \cite{Gogolin98a}. Consequently, the fermionic
quasiparticle excitations that are characteristic of Fermi liquids
become unstable in 1D conductors: instead, collective density
fluctuations constitute the stable elementary excitations in LLs.
Luttinger liquids are further characterized by power-law
correlations with interaction-dependent exponents, and by separation
of the spin and charge degrees of freedom. Power-law behaviors of
the differential conductance have been observed for edge states in
the fractional quantum Hall regime \cite{Chang96a} and metallic
single-walled carbon nanotubes (SWNTs)
\cite{Tans97a,Bockrath99a,Yao99a}. The spin-charge separation was
also observed in organic Bechgaard salts \cite{Lorenz02a}.

One-dimensional single-electron tunneling transistors (SETs) that
exhibit power-laws characteristic of Luttinger liquids have been
fabricated using semiconducting quantum wires \cite{Auslaender00a}
or metallic SWNTs \cite{Postma01a}. In an experiment using
semiconducting quantum wires, \citet{Auslaender00a} showed that the
widths of resonant levels of a 1D island embedded in an interacting
1D wire decrease as a power law over a range of temperatures, in a
quantitative agreement with the theoretical prediction by
\citet{Furusaki98a}. In an SWNT experiment, on the other hand,
\citet{Postma01a} studied a quantum dot (QD), formed by adjacent
defects in a long metallic SWNT, in a SET geometry. The conductance
as a function of temperature was seen to deviate from the
conventional predictions \cite{Kane92prb,Furusaki98a}. To explain
the unpredicted temperature dependence of the conductance at low
temperatures, \citet{Postma01a}, followed by \citet{Thorwart02a},
proposed a new transport mechanism, correlated sequential tunneling
(CST), in which additional quantum correlations due to Coulomb
interactions across the barriers were considered beyond the
conventional (uncorrelated) sequential tunneling (UST) approach.

The power-law exponent of the temperature dependence of conductance
in the UST and CST approaches in the strong interaction regime has
been studied by several groups. While the numerical approach using a
dynamical quantum Monte Carlo method supports CST approach
\cite{HugleS04epl}, the ``leading-log'' methods followed by the
functional renormalization group approaches does not support the CST
mechanism \cite{NazarovY03prl,PolyakovD03prb,MedenV04a,EnsT04a}.

Since the pioneering works by \citet{Kane92prb}, many properties
of the double barrier (DB) structure with the Luttinger liquid leads
have been investigated for quantum dots with single resonant level
\cite{Kane92b} and with many resonant levels \cite{Furusaki98a}.
Later, even the quantum dot was descried using the Luttinger model.
In this regime, new phenomena arise due to the interplay between
interactions within the 1D wire and the Coulomb blockade induced by
the confinement of electrons in a small region. Since the elementary
excitations in the system are plasmons (charge density waves), the
excitation spectrum of the QD is bosonic.

Very recently, various transport properties in such systems have
been studied by a number of  groups. Within the sequential tunneling
approach, Braggio \textit{et al.} found power-law-type differential
conductance with sharp peaks related to the activation of plasmons
\cite{Braggio00a}, charge-spin separation manifested in the
conductance peak positions \cite{Braggio01a}, and shot noise
indicating Luttinger liquid correlations \cite{Braggio03a}. The
authors consistently assumed fast relaxation of the plasmonic excitations
in the \textit{quantum dot}, implying that excitations
created by one tunneling event do not influence subsequent tunneling
events. We hereafter refer to this approach as ``equilibrium
plasmons''.

The present authors have focused more on the properties and the
consequences of the non-equillibrium distribution of plasmons in the
QD, in the following ``non-equilibrium plasmons''
\cite{KimJU03a,KimJU04a}. This is the case when the plasmon
excitations in the quantum dot redistribute only via the
single-electron tunneling events through tunnel barriers. We found
that while the steady-state plasmon distribution in the QD is highly
non-equilibrium, the average electric current is only weakly
affected by the non-equilibrium properties of plasmons
\cite{KimJU03a}. On the other hand, the non-equilibrium plasmons
\textit{do} affect more sensitive measurements such as shot noise
(SN) and full counting statistics (FCS). Both of these
characteristics show non-Poissonian behavior even at low bias
voltages: the shot noise is greatly enhanced above the Poissonian
limit (super-Poissonian) and the enhancement is more severe in the
strong interaction limit \cite{KimJU04a}.

 As an extension of our previous
work \cite{KimJU03a,KimJU04a}, we investigate the consequence of the
non-equilibrium plasmons on average current, shot noise, and full
counting statistics of a 1D-SET that consists of three Luttinger
liquid segments, based on a master equation approach in the
conventional sequential tunneling regime (ST).  In Sec.
\ref{sec:formalism} we introduce our model for a 1D quantum wire SET
and present the analytically obtained tunneling rates within the
golden rule approximation. In the same section, we also introduce
the master equation which is used to obtain all the results of this
work and discuss the possible experimental realizations. Sec.
\ref{sec:probability} is devoted to the discussion of the
distribution of the non-equilibrium occupation probabilities of
plasmonic many-body states. We then proceed to investigate the
consequence of the non-equilibrium plasmons in the context of the
transport properties. We first consider average current in Sec.
\ref{sec:current}, and then discuss shot noise in Sec.
\ref{sec:noise}. Finally, we investigate full counting statistics in
Sec. \ref{sec:FCS}. We conclude in Sec. \ref{sec:conclusions}.

\section{Formalism \label{sec:formalism}}

The electric transport of a double barrier structure in the
(incoherent) sequential tunneling regime can be described by the
master equation \cite{Kulik75a,Glazman89a,Furusaki98a}
\begin{widetext}
\begin{equation} \label{eq:MasterEquation}
\frac{\partial}{\partial t} P(N,\{n\},t)  =\sum_{N'}\sum_{\{n'\}}
\left[ \Gamma(N,\{n\} \leftarrow
N',\{n'\})P(N^\prime,\{n^\prime\},t) -\Gamma(N',\{n'\} \leftarrow
N,\{n\}) P(N,\{n\},t) \right],
\end{equation}
\end{widetext}
where $P(N,\{n\},t)$ is the probability that at time $t$ there are
$N$ (excess) electrons and $\{n\}=(n_1,n_2,\cdots,n_m,\cdots)$
plasmon excitations (i.e. collective charge excitations), that is,
$n_m$ plasmons in the mode $m$ on the quantum dot. The transitions
occur via single-electron tunneling through the left (L) or
right (R) junctions (see Fig. \ref{fig:ModelFig}). The total
transition rates $\Gamma$ in master equation \eqref{eq:MasterEquation}
are sums of the two transition rates $\Gamma_L$ and $\Gamma_R$
where $\Gamma_{L/R}(N,\{n\}\tol
N',\{n'\})$ is the transition rate from a quantum state
$(N',\{n'\})$ to another quantum state $(N,\{n\}$ via electron
tunneling through $L/R$-junction.

Master equation \eqref{eq:MasterEquation} implies that, with known
transition rates, the occupation probabilities $P(N,\{n\},t)$ can
be obtained by solving a set of linear first order differential equations
with the probability conservation
$\sum_{N,\{n\}}P(N,\{n\},t)=1$. In the long time limit
 the system converges to a steady-state with
probability distribution $\lim_{t\rightarrow
\infty} P(N,\{n\},t+\tau)=P_{st}(N,\{n\})$, irrespective of the
initial preparation of the system.

To calculate the transition rates, we start from the Hamiltonian
of the system.
The reservoir temperature is assumed zero ($T=0$), unless it
is stated explicitly.

\subsection{Model and Hamiltonian \label{subsec:Model}}

\begin{figure}[htb]
\includegraphics[width=8cm,height=5cm]{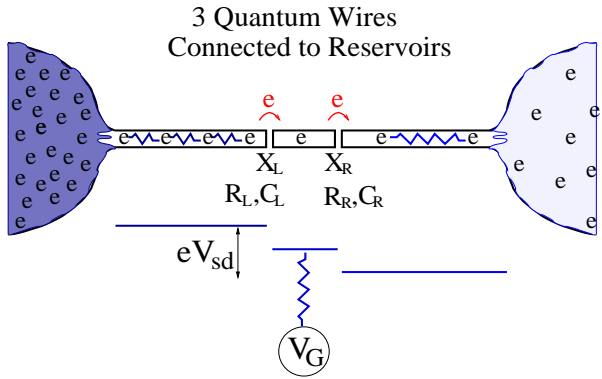}
\caption{(color online) Model system. Two long wires are adiabatically
  connected
to reservoirs and a short wire is weakly coupled to the two leads.
Tunneling resistances at junction points $X_L$ and $X_R$ are
$R_{L/R}$, and the junction capacitances are considered equal $C_L
= C_R$. Quantum dot is capacitively coupled to the gate
electrode.} \label{fig:ModelFig}
\end{figure}

The system we consider is a (1D) quantum wire SET. Schematic
description of the system is that a finite wire segment, which we
call a quantum dot, is weakly coupled to two long wires as depicted
in Fig. \ref{fig:ModelFig}. The chemical potential of the quantum
dot is controlled by the gate voltage ($V_G$) via a capacitively
coupled gate electrode. In the low-energy regime, physical
properties of the metallic conductors are well described by
linearized dispersion relations near the Fermi points, which allows
us to adopt the Tomonaga-Luttinger Hamiltonian for each wire
segment. We model the system with two semi-infinite LL leads and a
finite LL for the central segment. The leads are adiabatically
connected to reservoirs which keep them in internal equilibria. The
chemical potentials of the leads are controlled by source-drain
voltage ($V$), and the wires are weakly coupled so that the
single-electron tunneling is the dominant charge transport
mechanism, i.e. we are interested in the sequential tunneling
regime. Rigorously speaking, the voltage drop between the two leads
($V$) deviates from the voltage drop between the left and right
reservoirs (say $U$) if electron transport is activated
\cite{Egger96a,Egger98a}. However, as long as the tunneling
amplitudes through the junctions (barriers) are weak so that the
Fermi golden rule approach is appropriate, we estimate $V\approx U$.

The total Hamiltonian of the system is then given by the sum of the
bosonized LL Hamiltonian $\hat{H}_0=\hat H_L+\hat H_D + \hat H_R$
accounting for three isolated wire segments labeled by
$\ell=(L,D,R)$, and the tunneling Hamiltonian
$\hat{H}_T$ accounting for single-electron hops through the
junctions L and R at $X_L$ and $X_R$ respectively,
\begin{equation}
\hat{H}=\hat{H}_{0} + \hat H_T.
\label{eq:H}
\end{equation}

Using standard bosonization technique, the Hamiltonian describing
each wire segment can be expressed in terms of  creation
and annihilation operators for collective excitations
($\hat b^{\dagger}$ and $\hat b$)
\cite{Voit94rpp,Gogolin98a}. For the semi-infinite leads, it reads
\begin{equation}\label{eq:H_L/R}
\hat H_{\ell} = \sum_{\nu=1}^M
\veps^{(\ell)}_{\nu} \sum_{m=1}^{\infty} m
\hat b^{\dagger}_{\nu,m}\hat b_{\nu,m}~, ~\mbox{ for } ~\ell=L,R~\,,
\end{equation}
where the index $\nu$ labels the $M$ transport sectors of the conductor
and $m$ the wave-like collective excitations on each transport sector.
The effects of the Coulomb interaction in 1D wire are characterized by
the Luttinger parameter $g_\nu$:
$g=1$ for noninteracting Fermi gas and $0<g<1$ for the repulsive
interactions ($g\ll 1$ in the strong interaction limit). The Coulomb
interaction also renormalizes the Fermi velocity to
$v_\nu=v_F/g_\nu$.  The energy of an elementary excitation in sector
$\nu$ is given by $\veps_{\nu}=\pi\hbar v_\nu/L$ where $L$ is the
length of the wire and $\hbar$ the Planck constant. For instance, if
the wire has a single transport channel (usually referred to as
spinless electrons), e.g. a wire with one transport channel under a
strong magnetic field, the system's dynamics is determined by
collective charge excitations (plasmons) alone ($\nu=\rho$ and
$M=1$).  If, however, the spin degrees of freedom survive, the wire
has two transport sectors ($M=2$); plasmons ($\nu=\rho$) and
spin-waves ($\nu=\sigma$) \cite{Voit94rpp}. If the system has two
transport channels with electrons carrying spin ($M=4$), as is the
case with SWNTs, the transport sectors are total-charge-plasmons
($\nu=\rho$), relative-charge-plasmons ($\nu=\Delta \rho$),
total-spin-waves ($\nu=\sigma$), and relative-spin-waves
($\nu=\Delta\sigma$) \cite{Kane97a,Egger97a}.

For the short central segment, the zero-mode need to be accounted
for as well, which yields
\begin{multline} \label{eq:H_D}
\hat H_D= \sum_{\nu}\veps_{\nu} \sum_{m=1}^{\infty} m \hat
b^{\dagger}_{\nu,m}\hat b_{\nu,m} \\
+\frac{\veps_{\rho}}{2Mg_{\rho}}(\hat N_{\rho}-N_G)^2
+\sum_{\nu\neq \rho}\frac{\veps_{\nu}}{2Mg_{\nu}} \hat
N_{\nu}^2-E_r.
\end{multline}

In the second line of Eq. \eqref{eq:H_D}, which represents the
zero-mode energy of the quantum dot, the operator $\hat{N_\nu}$
measures the ground state charge, i.e. with no
excitations, in the $\nu$-sector. The zero-mode energy
systematically incorporates Coulomb interaction in terms of the
Luttinger parameter $g_\rho$ in the QD.  To refer to the zero-mode
energy later in this paper, we define the
``charging energy'' $E_C$ as the minimum energy cost to add an excess
electron to the QD in
the off-Coulomb blockade regime, i.e.,
\begin{equation} \label{eq:E_C}
E_C = [E_D(2,0)-E_D(1,0)]_{N_G=1/2}=\veps_p/g \,.
\end{equation}
Note that this is twice the conventional definition.  The charging
energy vanishes in the noninteracting limit ($g_\rho=1$) and becomes the
governing energy scale in the strong interaction limit ($g_\rho \ll 1$).
The
origin of the charging energy in conventional quantum dots is the
long range nature of the Coulomb interaction. In the theory of
Luttinger liquid, the long range interaction can easily be
incorporated microscopically through the interaction strength
$g_\rho$. For the effect of the finite-range interaction across a
tunneling junction, for instance see Refs.
\onlinecite{SassettiM97ssc,SassettiM97prb}.

Note that charge and spin are decoupled in Luttinger liquids, which
implies the electric forces affect the (total) charge sector only;
due to intrinsic $e-e$ interactions, $g_\rho < 1$ but
$g_{\nu\neq\rho} = 1$, and the gate voltage shifts the band bottom
of the (total) charge sector as seen by the dimensionless gate
voltage parameter $N_G$ in Eq. \eqref{eq:H_D}. As will be shown
shortly, the transport properties of the L/R--leads are determined
by the LL interaction parameter $g_\rho$ and the number of the
transport sector $M$. In this work, we consider each wire segment
has the same interaction strength for the (total) charge sector, $g
\equiv g^{(L)}_{\rho}=g^{(R)}_{\rho}=g^{(D)}_{\rho}$. Accordingly,
the energy scales in the quantum dot are written by $\veps_\rho
\equiv \veps^{(D)}_{\rho} = \pi\hbar v_F/g_\rho L_D$ and $\veps_0
\equiv \veps_{\nu\neq\rho}=\pi\hbar v_F/L_D$.

 We consider the ground state energy in the QD is the same as
 those in the leads, by choosing the reference energy $E_r$ in
Eq. \eqref{eq:H_D} equals the minimum value of the zero-mode
energy,
\begin{equation}\label{eq:E_r}
E_r = \min\left(\frac{\veps_\rho N_G^2}{2Mg_\rho},
\frac{\veps_\rho (1-N_G)^2}{2Mg_\rho}+\frac{\veps_0
(M-1)}{2M}\right),
\end{equation}
where $\min(x,x^\prime)$ denotes the smaller of $x$ and
$x^\prime$, and the gate charge $N_G$ is in the range $N_G \in
[0,M]$.

The zero-mode energy in the QD,
\begin{equation} \label{eq:E_0}
E_0 =\frac{\veps_\rho}{2Mg_\rho}(N-N_G)^2
+\frac{\veps_0}{2M}\sum_{\nu\neq \rho} N_{\nu}^2-E_{r}
\end{equation}
yields degenerate ground states for $N=0$ and $N=1$ excess electrons
when $N_G=\big[ (M-1)g_\rho^2+1\big]/2$. Here we replaced $N_\rho$
by the number of the total excess electrons $N$ since $N_{\rho}=N$,
and $N_{\nu}$ are all either even or odd integers, simultaneously;
in the case of the SWNTs with $N=\sum_{i,s} N_{i,s}$ excess
electrons, where $i=1,2$ is the channel index and
$s=\uparrow,\downarrow$ is the spin index of conduction electrons
(M=4), $N_{\rho}=N$, $N_{\sigma}=\sum_{i}
(N_{i,\uparrow}-N_{i,\downarrow})$,
$N_{\Delta\rho}=\sum_{s}(N_{1,s}-N_{2,s})$, and $N_{\Delta\sigma}=
\sum_{i}(-1)^{i} (N_{i,\uparrow}-N_{i,\downarrow})$.

From now on we consider only one spin-polarized (or spinless)
channel unless otherwise stated --- our focus is on the role of
Coulomb interactions, and the additional channels only lead to more
complicated excitation spectra without any qualitative change in the
physics we address below.  A physical realization of the single-channel case
may be obtained {\em e.g} by exposing the quantum wire to a large
magnetic field.

For the system with high tunneling barriers, the electron
transport is determined by the bare electron hops at the tunneling
barriers. The tunneling events in the DB structure are described by
the Hamiltonian
\begin{equation} \label{eq:H_T}
 \hat H_T =  \sum_{\ell=L,R}
\left[ t_\ell \hat \Psi^\dagger _{D} (X_\ell)\hat \Psi_{\ell}
(X_\ell) +H.C. \right],
\end{equation}
where $\hat\Psi^\dagger_{l}(X_\ell)$ and $\hat\Psi_{l}(X_\ell)$ are
the electron creation and annihilation operators at the edges of the
wires near the junctions at $X_L$ and $X_R$. As mentioned earlier,
the electron field operators $\hat \Psi$ and $\hat \Psi^\dagger$ are
related to the plasmon creation and annihilation operators $b$ and
$b^\dagger$ by the standard bosonization technique. Different
boundary conditions yield different relations between electron field
operators and plasmon operators. Exact solutions for the periodic
boundary condition have been known for decades
\cite{Haldane81b,Voit94rpp} but the open boundary conditions which
are apt for our system of consideration has been investigated only
recently (See for example Refs.
\onlinecite{FabrizioM95prb,Eggert96a,Kane97a,Mattsson97a}).

The dc bias voltage $V=V_L+V_R$ between L and R leads is
incorporated into the phase factor of the tunneling matrix
elements $t_\ell=|t_\ell|\exp(\mp i eV_\ell t/\hbar)$ by a
time-dependent unitary transformation \cite{Ingold92a}. Here
$V_{L/R}= V C/C_{R/L}$ is voltage drop across the L/R--junction
where $C=C_LC_R/(C_L+C_R)$ is the effective total capacitance of
the double junction, and the bare tunneling matrix amplitudes
$|t_{L/R}|$ are assumed to be energy independent.
Experimentally, the tunneling matrix amplitude is sensitive to the
junction properties while the capacitance is not.  For simplicity, the
capacitances are thus assumed to be symmetric $C_L=C_R$ throughout this
work.  By junction asymmetry we mean the asymmetry in (bare) squared
tunneling amplitudes $|t_{L/R}|^2$. The parameter $R= |t_L|^2/|t_R|^2$
is used to describe junction asymmetry; $R=1$ for symmetric junctions
and $R\gg 1$ for a highly asymmetric junctions.

It is known that, at low energy scales in the quantum wires with
the electron density away from half-filling, the processes of
backward and Umklapp scattering, whose processes generate momentum
transfer across the Fermi sea ($\approx 2k_F$), can be safely
ignored in the middle of ideal 1D conductors \cite{Voit94rpp},
including armchair SWNTs \cite{Kane97a}. The Hamiltonian
\eqref{eq:H} does not include the backward and Umklapp scattering
(except at the tunneling barriers) and therefore it is valid away
from half-electron-filling.

We find that, in the regime where electron spin does not play a role,
the addition of a transport channel does not change essential
physics present in a single transport channel. Therefore, we
primarily focus attention to a QW of single transport channel with
spinless electrons and will comment on the effects due to multiple
channel generalization, if needed.

\subsection{Electron transition rates \label{subsec:T_rates}}

The occupation probability of the quantum states in the SET system
changes via electron tunneling events across L/R--junctions. In the
single-electron tunneling regime, the bare tunneling amplitudes
$|t_{L/R}|$ are small compared to the characteristic energy scales
of the system and the electron tunneling is the source of small
perturbation of three isolated LLs. In this regime, we calculate
transition rates $\Gamma_{L/R}$ between eigenstates of the
unperturbed Hamiltonian $H_0$ to the lowest non-vanishing order in
the tunneling amplitudes $|t_{L/R}|$. In this golden rule
approximation, we integrate out lead degrees of freedom since the
leads are in internal equilibria, and the transition rates are given
as a function of the state variables and the energies of the QD only
\cite{KimJU03a,Braggio00a},
\begin{multline}\label{eq:G_L/R}
\Gamma_{L/R}\left(N',\{n'\}\leftarrow N,\{n\}\right)  \\ =
\frac{2\pi}{\hbar}|t_{L/R}|^2\, \gamma(W_{L/R})\,
\gamma_D(\{n'\},\{n\}) \, .
\end{multline}
 In Eq. \eqref{eq:G_L/R} $W_{L/R}$ is the change in the
Gibbs free energy associated with the tunneling across the
L/R--junction,
\begin{multline}\label{eq:W_L/R}
W_{L/R} = E_D(N',\{n'\}) - E_D(N,\{n\}) \\\mbox{}%
\mp (N'-N) eV_{L/R} \, .
\end{multline}
Here $E_D(N,\{n\})=\langle N,\{n\}|\hat H_D|N,\{n\}\rangle $ is
the energy of the eigenstate $|N,\{n\}\rangle$ of the dot and $L/R$
correspond to $-/+$. For the QD with only one transport channel
with spinless electrons only,
\begin{equation}
\label{eq:ED1} E_D (N,\{n\}) =\veps_p \left[
  \sum_{m=1}^\infty m n_{m}  +
  \frac{(N-N_G)^2}{2g_D}\right]-E_r\, ,
\end{equation}
where $\veps_p=\veps_\rho$ accounting that we consider only charge
plasmons and $n_{m}=\langle n_m| \hat b^{\dagger}_m \hat b_{m}|n_m
\rangle$ is the number of plasmons in the mode $m$.

The function $\gamma(x)$ in Eq. \eqref{eq:G_L/R} is responsible
for the plasmon excitations on the leads, and given by (see e.g.
Ref. \onlinecite{Furusaki98a})
\begin{widetext}
\begin{equation}
\label{eq:gamma} \gamma(\veps)
=\frac{1}{2\pi\hbar}\int_{-\infty}^{\infty} dt e^{i\veps t}\langle
\Psi_{\ell}(X_\ell,0) \Psi_{\ell}^\dagger (X_\ell,t)\rangle
=\frac{1}{2\pi\hbar}\frac{1}{\pi v_F} \left(\frac{2\pi\Lambda
g^{1/(1-g)}}{\hbar v_F\beta}\right)^\alpha
\left|\Gamma\left(\frac{1+\alpha}{2}+i\frac{\beta\veps}{2\pi}\right)
\right|^2 \frac{e^{-\beta\veps/2}}{\Gamma(1+\alpha)} \,,
\end{equation}
\end{widetext}
where $\beta=1/k_B T$ is the inverse temperature in the leads,
$\Lambda$ is a short wavelength cutoff, and $\Gamma(z)$ is the
Gamma function. The exponent $\alpha=(g^{-1}-1)/M$ is a
characteristic power law exponent indicating interaction strength
of the leads with $M$ transport sectors (hence, in our case,
$M=1$). At $g=1$ (non-interacting case), the exponent $\alpha=0$
and it grows as $g\to 0$ (strong interaction). The decrease of the
exponent $\alpha$  with increasing $M$ implies that the effective
interaction decreases due to multi-channel effect, and the
Luttinger liquid eventually crosses over to a Fermi liquid in the
many transport channel limit \cite{MatveevK93prl,MatveevK93phyb}.

For the
non-interacting electron gas, the spectral density $\gamma(\veps)$
is the TDOS multiplied by the Fermi-Dirac distribution function
$f_{FD}(\veps)=[1+\exp(\beta \veps)]^{-1}$;
$\gamma(\veps)=\frac{1}{\pi\hbar v_F} f_{FD} (\veps)$ for $g=1$.
At zero temperature, $\gamma(\veps)$ is proportional to a power of
energy,
\begin{equation}
\lim_{T\rightarrow 0}\gamma(\veps) = \Theta(-\veps)
\frac{1}{\pi\hbar v_F}
\frac{\left(|\veps|/\veps_\Lambda\right)^\alpha}{\Gamma(\alpha+1)},
\label{eq:gamma_T0}
\end{equation}
where $\veps_\Lambda = \hbar v_F/\Lambda g^{1/(1-g)}$ is a high
energy cut-off. At zero temperature $\gamma(\veps)$
 is the TDOS for the negative energies and zero otherwise (as it should be),
imposed by the unit step function $\Theta(-\veps)$.

 The function $\gamma_D$ in Eq. \eqref{eq:G_L/R}
 accounts for the plasmon  transition amplitudes in the QD,
and is given by
 \begin{multline} \label{eq:gamma_D}
 \gamma_D(\{n'\},\{n\}) =
 \delta_{N',N+1} |\langle  N',\{n'\}|\hat\Psi^\dagger_{D}(X_\ell)
 |N,\{n\}\rangle|^2 \\
 +  \delta_{N',N-1}|\langle  N',\{n'\}| \hat\Psi_{D}(X_\ell)
 |N,\{n\}\rangle|^2
 \end{multline}
where we used that the
 zero-mode overlap is unity for $N^\prime=N\pm 1$ and vanishes
 otherwise.  The overlap integrals between plasmon modes are, although
 straightforward, quite tedious to calculate, and we refer
 to Appendix \ref{app:OndotT} for the details. The resulting overlap of the
 plasmon states can be written as a function of the mode occupations
 $n_m$,
\begin{multline}
\label{eq:Laguerre} |\langle
\{n^\prime\}|\hat\Psi^\dagger_{D}(X_\ell)|\{n\}\rangle|^2 =|\langle
\{n^\prime\}|\hat\Psi_{D}(X_\ell)|\{n\}\rangle|^2
\\ =\frac{1}{L_D}\left(\frac{\pi \Lambda}{L_D}\right)^{\alpha_D}
\phi(\{n'\},\{n\})
\end{multline}
with
\begin{multline} \label{eq:phi_nu}
\phi(\{n'\},\{n\}) \\
=\prod_{m=1}^{\infty} \left(\frac{1}{g m}\right)^{|n'_m-n_m|}
\frac{n_m^{(<)} !}{n^{(>)}_m !}
\left[L^{|n_m'-n_m|}_{n_m^{(<)}}\left(\frac{1}{g
m}\right)\right]^2
\end{multline}
where $n_m^{(<)}=\min(n_m^\prime,n_m)$ and $n_m^{(>)} =
\max(n_m^\prime,n_m)$,
 $\alpha_D=(g_D^{-1}-1)$ for the QD with one transport sector, and
 $L_{a}^{b}(x)$ are Laguerre polynomials. Additional transport
 sectors would appear as multiplicative factors of the same form as
 $\phi(\{n'\},\{n\})$ and result in a reduction of the exponent
 $\alpha_D$.
Notice that in the low energy scale only the first few occupations
$n_m$ and $n'_m$ in the product differ from zero, participating to
the transition rate \eqref{eq:G_L/R} with nontrivial
contributions.

\subsection{Plasmon relaxation process in the quantum dot
 \label{subsec:RelaxationRate}}

In general, plasmons on the dot are excited by
tunneling events and have a highly non-equilibrium distribution. The
coupling of the system to the environment such as external circuit
or background charge in the substrate leads to relaxation towards
the equilibrium. While the precise form of the relaxation rate,
$\Gamma_p$, depends on the details of the relaxation mechanism, the
physical properties of our concern do not depend on the details.
Here we take a phenomenological model where the plasmons are
coupled to a bath of harmonic oscillators by
\begin{equation}
\label{Noise-llj-long::eq:bath}
H_\mathrm{plasmon-bath} = \sum_{m,n}\sum_\alpha
\left(g_{mn}^\alpha\, b_m^\dag b_n a_\alpha + h.c.\right) \,.
\end{equation}
In Eq.~(\ref{Noise-llj-long::eq:bath}) $a_\alpha$ and $a_\alpha^\dag$
are bosonic operators describing the oscillator bath and $g_{mn}^\alpha$
is the coupling constants.  We will assume an Ohmic form of the bath
spectral density function
\begin{equation}
\label{Noise-llj-long::eq:bath2} J_{mn}(\omega) \equiv \sum_\alpha
|g_{mn}^\alpha|^2\,\delta(\omega-\omega_\alpha) =
\gamma_p\Gamma_0\omega \,,
\end{equation}
where $\omega_\alpha$ is the frequency of the oscillator
corresponding to $a_\alpha$, $\gamma_p$ is a dimensionless constant
characterizing the bath spectral density, and
$\Gamma_0^{-1}=\hbar^2 v_FL_D(|t_L|^{-2} + |t_R|^{-2})$ is the natural
time scale of the system.
Within the rotating-wave approximation, the plasmon transition rate
due to the harmonic oscillator bath is given by
\begin{equation}
\label{eq:Gp} \Gamma_p(\{n'\}\tol \{n\}) = \gamma_p\Gamma_0
\frac{W_p/\veps_p}{e^{\beta W_p} - 1}
\end{equation}
with
\begin{math}
W_p = \veps_p\sum_m(n_m'-n_m)
\end{math},
where $\veps_p = \veps_\rho$ is the plasmon energy.
Note that these
phenomenological rates obey detailed balance and, therefore, at low
temperatures only processes that reduce the total plasmon energy
occur with appreciable rates.

\subsection{Matrix formulation \label{subsec:MatrixNotation}}

For later convenience, we introduce a matrix notation for the
transition rates $\Gamma$, with the matrix elements defined by
\begin{equation}
\label{eq:G_mat_1}
\left[\whatbf\Gamma_\ell^\pm(N)\right]_{\{n^\prime\},\{n\}} =
\Gamma_\ell(N\pm 1,\set{n'}\tol N,\set{n}) \,,
\end{equation}
i.e, the element $(\{n^\prime\},\{n\})$ of the matrix block
$\whatbf\Gamma_\ell^\pm(N)$ is the transition rate
$\Gamma_\ell(N\pm 1,\set{n'}\tol N,\set{n})$. Similarly,
\begin{equation}
\label{eq:G_mat_2}
\left[\whatbf\Gamma_\ell^0\right]_{\{n'\},\{n\}} \\\mbox{}%
= \delta_{\{n'\},\{n\}} \sum_{\{n''\}} \left[\whatbf\Gamma_\ell^+
  + \whatbf\Gamma_\ell^-\right]_{\{n''\},\{n\}} \,,
\end{equation}
and
\begin{multline}
\label{eq:G_mat_3} \left[\whatbf\Gamma_p(N)\right]_{\{n'\},\{n\}}
= - \Gamma_p(\set{n'},\set{n}) \\\mbox{}%
+ \delta_{\{n'\},\{n\}}\sum_{\{n''\}} \Gamma_p(\set{n''},\set{n})
\,.
\end{multline}
Master equation \eqref{eq:MasterEquation}
can now be conveniently expressed as
\begin{equation}
\label{eq:ME_mat} \frac{d}{dt}\ket{P(t)} =
-\whatbf\Gamma\ket{P(t)}
\end{equation}
with
\begin{math}
\whatbf\Gamma = \whatbf\Gamma_p +
\sum_{\ell=L,R}\left(\whatbf\Gamma_\ell^0 - \whatbf\Gamma_\ell^+
  - \whatbf\Gamma_\ell^-\right)
\end{math},
where $\ket{P(t)}$ is the column vector (not to be confused with
the ``ket'' in quantum mechanics) with elements given by
\begin{math}
\braket{N,\set{n}|P(t)} \equiv P(N,\set{n},t)
\end{math}.
Therefore, the time evolution of the probability vector satisfies
\begin{equation}
\label{eq:Pt_vec} \ket{P(t)} = \exp(-\whatbf\Gamma t) \ket{P(0)}.
\end{equation}
In the long time limit, the system reaches a steady state
$\ket{P(\infty)}$.

The ensemble averages of the matrices $\whatbf{\Gamma}^\pm_{\ell}$
can then be defined by
\begin{equation} \label{eq:Gamma_avg}
\avg{\whatbf{\Gamma}^\pm_{\ell}(t)}=\sum_{N,\{n\}}\bra{N,\{n\}}
\whatbf{\Gamma}^\pm_{\ell}\ket{P(t)}\, .
\end{equation}
We will construct other statistical quantities such as average
current and noise power density based on Eq. \eqref{eq:Gamma_avg}.

\section{Steady-state probability distribution of nonequilibrium plasmons
 \label{sec:probability}}

By solving the master equation \eqref{eq:ME_mat} numerically
(without plasmon relaxation), in Ref. \onlinecite{KimJU03a}, we
obtained the occupation probabilities of the plasmonic many-body
excitations as a function of the bias voltage and the interaction
strength. We found that in the weak to noninteracting regime,
$\alpha\approx 0$ or $g=1/(1+\alpha) \approx 1$ for the wire with
one transport sector, the non-equilibrium probability of plasmon
excitations is a complicated function of the detailed configuration
of state occupations $\{n\}=(n_1,n_2,\cdots,n_m,\cdots)$.

In contrast, the non-equilibrium occupation probability in the
strong interaction regime with (nearly) symmetric tunneling barriers
depends only on the total energy of the states, and follows a
universal form \emph{irrespective of electron charge $N$} in the QD.
In the leading order approximation, it is given by
\begin{equation}
P^{(0)}(\veps) \approx \frac{1}{Z}\exp\left(-\frac{3(\alpha+1)}{2}
\frac{\veps}{eV}\log\frac{\veps}{\veps_p}\right),
\label{eq:P_univ}
\end{equation}
where $Z$ is a normalization constant. Notice that $\veps$ is the
total energy, including zero-mode and plasmon contributions. The
distribution has a universal form which depends on the bias voltage
and the interaction strength. The detailed derivation is in appendix
\ref{app:P_univ}. This analytic form is valid for the not too low
energies $(\veps, eV) \gtrsim 3 \veps_p$ and in the strongly
interacting regime $\alpha \gtrsim 1$. More accurate approximation
formula (Eq. \eqref{eq:app_univ_P1}) is  derived in appendix
\ref{app:P_univ}.

\begin{figure}[htbp]
\includegraphics[width=8cm]{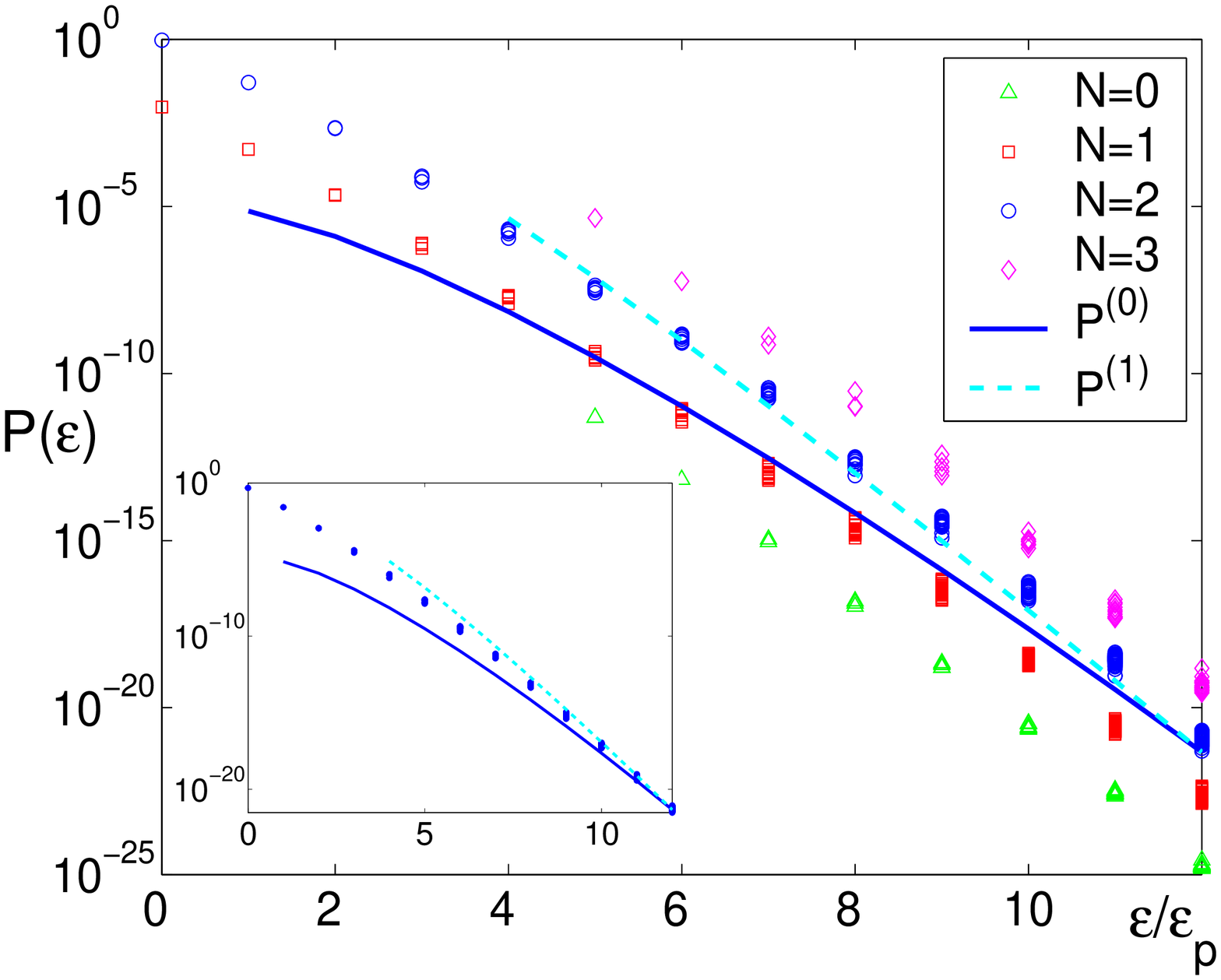}\\ (a) \\
\includegraphics[width=8cm]{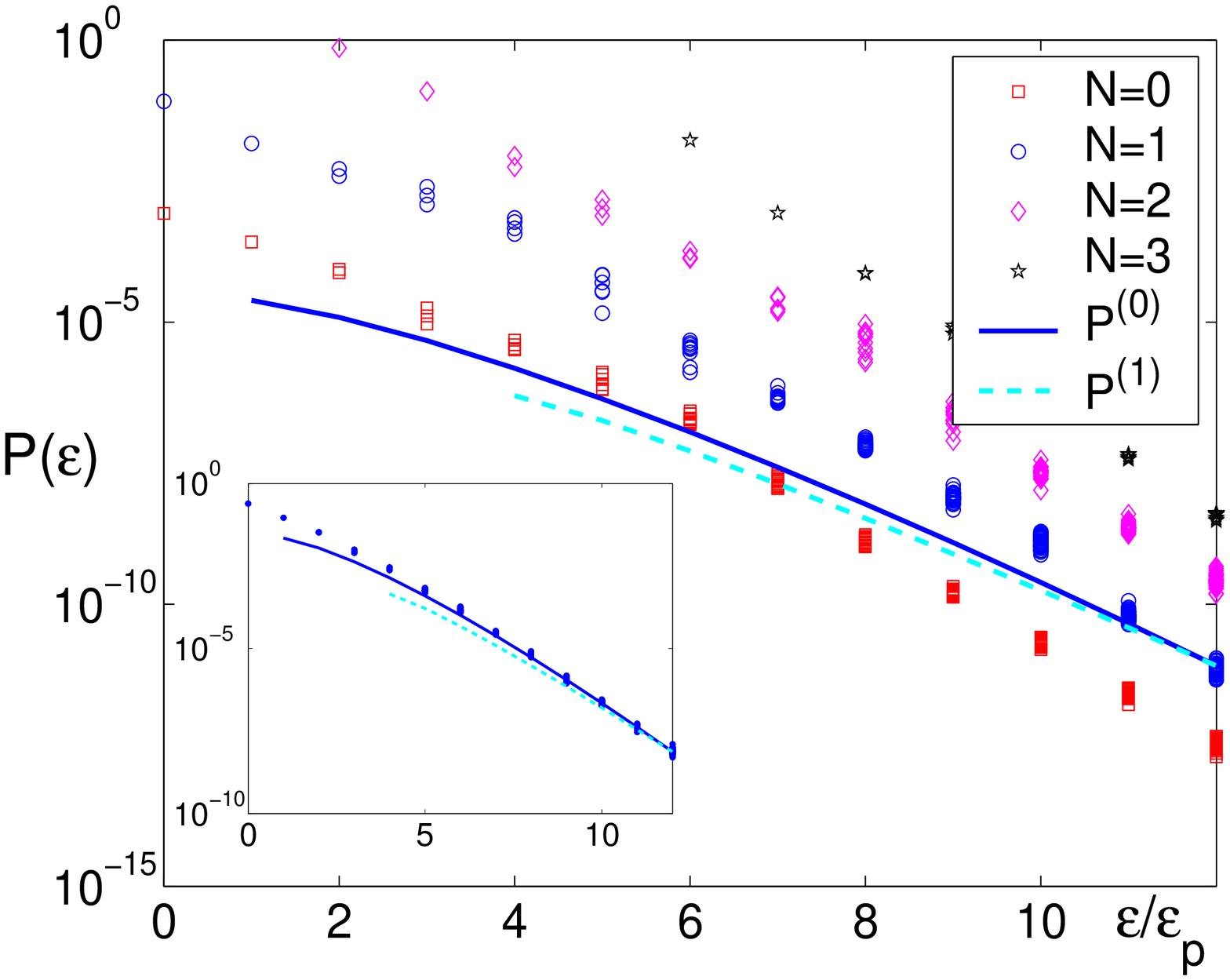}\\ (b)
\caption{(color online) The occupation probability $P(\veps)$ as a function of
the mode energy $\veps/\veps_p$. The energy $\veps$ is
abbreviation for $E_D(N,\{n\})$, the bias $eV=6\veps_p$, the
asymmetry parameter $R=100$, and $N_G=1/2$ ($T=0$).  Two analytic
approximations, Eq. \eqref{eq:P_univ} (blue curve) and Eq.
\eqref{eq:app_univ_P1} (cyan curve), are fitted to the probability
distribution of the charge mode $N=1$ (blue circle). The
interaction parameter is (a) $g=0.2$ and (b) $g=0.5$. In the inset
the case of symmetric junctions ($R=1$) is plotted with the same
conditions.} \label{fig:P_univ}
\end{figure}

For symmetric junctions, the occupation probabilities fall on a
single curve, well approximated by the analytic formulas Eqs.
\eqref{eq:P_univ} and \eqref{eq:app_univ_P1}, as seen in the insets
in Fig. \ref{fig:P_univ}, where $P(\veps)$ is depicted as a function
of the state energies for (a) $g=0.2$ and (b) $g=0.5$, with
parameters $R=1$ for the inset and $R=100$ for the main figures ($eV
= 6\varepsilon_p$, $N_G=1/2$). For the asymmetric junctions, the
line splits into several branches, one for each electric charge $N$,
see the figure. However, as seen in \ref{fig:P_univ}(a), if the
interaction is strong enough ($g \lesssim 0.3$ for $R=100$ and $eV =
6\veps_p$), each branch is, independently, well described by Eq.
\eqref{eq:P_univ} or \eqref{eq:app_univ_P1}. For weaker
interactions, $g \agt 0.3$ for $R=100$ and $eV = 6\varepsilon_p$,
the analytic approximation is considerably less accurate as shown in
\ref{fig:P_univ}(b). Even in the case of weaker interactions,
however, the logarithms of the plasmon occupation probabilities
continue to be nearly linear in $\veps$ but with a slope that
deviates from that seen for symmetric junctions.

\section{Average current \label{sec:current}}

In terms of the tunneling current matrices $\hatbf{I}_{L/R}$
across the junction $L/R$
\begin{equation} \label{eq:I_mat_L/R}
\hatbf{I}_{L/R} = \mp e\left(\whatbf\Gamma_{L/R}^+ -
\whatbf\Gamma_{L/R}^-\right),
\end{equation}
the average current $I_{L/R}(t)=\avg{\hatbf I_{L/R}(t)}$ through
$L/R$--junction is
\begin{equation}\label{eq:I_avg_L/R}
I_{L/R}(t) = \sum_{N,\{n\}}\bra{N,\{n\}} \hatbf{I}_{L/R}
\ket{P(t)}.
\end{equation}
The total external current $I(t)=\avg{\hatbf I(t)}$, which includes
the displacement currents associated with charging and discharging
the capacitors at the left and right tunnel junctions, is then
conveniently written as
\begin{equation}\label{eq:I_avg}
I(t) = \sum_{\ell=L,R}(C/C_\ell) I_\ell (t),
\end{equation}
where
$C^{-1} = C_L^{-1} + C_R^{-1}$. As the system reaches steady-state in
the long time limit,
the charge current is conserved throughout the system,
$I=I(\infty)=I_L(\infty)=I_R(\infty)$.

\begin{figure}[htbp]
\includegraphics[width=8cm]{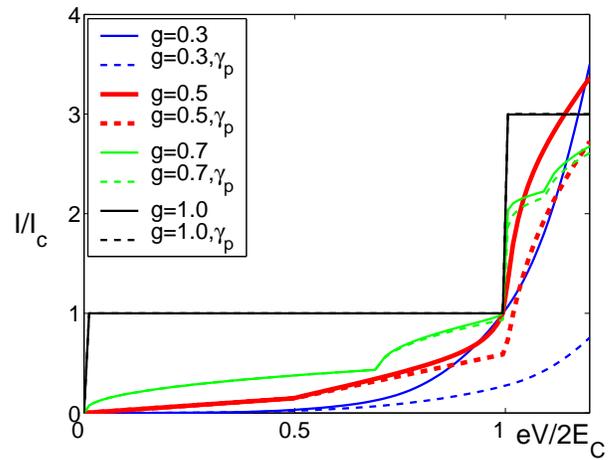}
\caption{(color online) Average current $I/I_c$ as a function of the
  bias voltage
$eV/2E_C$ and LL interaction parameter $g$ (black line for $g=1$,
green $g=0.7$, red $g=0.5$, blue $g=0.3$, respectively) for
$R=100$ (highly asymmetric junctions) with no plasmon relaxation
($\gamma_p=0$, solid lines) or with fast plasmon relaxation
($\gamma_0=10^4$, dashed lines). The bias voltage is normalized by
the charging energy $2E_C$ and current is normalized by the
current at $eV=2E_C$ with no plasmon relaxation for each $g$.
Other parameters are $N_G=1/2$, $T=0$.} \label{fig:IV_R100}
\end{figure}

One  consequence of non-equilibrium plasmons is the increase in
current as shown in Fig. \ref{fig:IV_R100}, where the average
current is shown as a function of the bias for different interaction
strengths. The currents are normalized by $I_c=I(eV=2E_C)$ with no
plasmon relaxation ($\gamma_p=0$) for each interaction strength $g$,
and we see that the current enhancement is substantial in the strong
interaction regime ($g\lesssim 0.5$), while there is effectively no
enhancement in noninteracting limit $g=1$ (the two black lines are
indistinguishable in the figure). In the weak interaction limit the
current increases in discrete steps as new transport channels become
energetically allowed, while at stronger interactions the steps are
smeared to power laws with exponents that depend on the number of
the plasmon states involved in the transport processes.

 Including the spin sector results in additional peaks in the
average current voltage characteristic that can be controlled by
the transverse magnetic field \cite{Braggio01a,CavaliereF04prl}.

The current-voltage characteristics show that, in the
non-interaction limit, the non-equilibrium approach predicts similar
behavior for the average current as the detailed balance approach
which assumes thermal equilibrium in the QD. In contrast, in the
strong interaction regime, non-equilibrium effects give rise to an
enhancement of the particle current.

Experimentally, however, the current enhancement may be difficult to
attribute to plasmon distribution as the current levels depend on
barrier transparencies and plasmon relaxation rates, and neither of
them can be easily tuned. We now turn to another experimental probe,
the shot noise, which is more sensitive to non-equilibrium effects.

\section{Current noise \label{sec:noise}}

Noise in electronic conductors is given by the ensemble average of
the current-current correlations \cite{Kogan96a,Beenakker03a}.
Thermal fluctuations and the discrete nature of the electron charge
are two fundamental sources of the noise; in specific devices there
may additional noise sources due to, e.g., fluctuating environmental
variables. Thermal (equilibrium) noise is not very informative since
it does not provide more information than the equilibrium
conductance of the system. In contrast, shot noise, which is a
consequence of the discreteness of charge and the stochastic nature
of transport, can provide further insight beyond average
current since it is a sensitive function of the correlation
mechanism, internal excitations, and the statistics of the
charge carriers \cite{deJong97a,Blanter00a}.

Influence of quantum coherence on shot noise is an intriguing issue.
It is known that the ensemble averaged quantum mechanical
calculations of shot noise to the leading order is identical to the
semiclassical approaches when the comparable theories are available
\cite[in Sec. 5]{Blanter00a}. However, the word ``semiclassical''
should not be confused with the deterministic motion of the
transport charges. For instance, \citet{Oberholzer02a} discuss the
crossover from full quantum to classical shot noise, by tuning the
electron dwell time in chaotic cavities, where by classical it means
the deterministic nature of electron motion.

Shot noise in interacting one-dimensional systems has also been the
subject of many recent works. For instance, the shot noise of the
edge states was used to measure the factional charge $ge$ of the
quasiparticles in the fractional quantum Hall
states~\cite{Kane94a,SaminadayarL97prl,Picciotto97a,ComfortiE02nature}.

 The shot noise
of double-barrier structures was widely studied in the last decade.
In conventional SET structures, shot noise is suppressed below
the Poisson limit due to the Coulomb correlations (in addition to
the Fermi correlation): the $e-e$ correlations typically result in
reduction of shot noise. Both quantum mechanical approaches
\cite{Chen91a,LundBo96a} and semiclassical derivations based on a
master equation approach predict identical shot noise results
\cite{Davies92a,Chen92a,Chen93a}, implying that the shot noise is
not sensitive to the quantum coherence in DB structures. If the
leads are superconducting in the SET structure, the tunneling
particles are either single-electrons or Cooper-pairs and the shot
noise is a functional of the dephasing process of the Cooper-pairs
\cite{Deblock03a,ChoiMS01c,ChoiMS03a}.

Semiclassical theories of shot noise based on a master equation
approach in the sequential tunneling regime for a SET have been
developed by many authors
\cite{Davies92a,Hershfield93a,Chen92a,Korotkov92a,Korotkov94a,Hanke94a}.
The predictions of some of these theories \cite{Hershfield93a} have
been experimentally confirmed \cite{Birk95a}.

The noise power density in the steady state is given by
\begin{equation}
\label{eq:Spower} S(\omega) =
\lim_{t\to\infty}2\int_{-\infty}^\infty{d\tau}\;
e^{+i\omega\tau}\left[
  \avg{\hatbf{I}(t+\tau)\hatbf{I}(t)} - \avg{\hatbf{I}(t)}^2
\right] \, .
\end{equation}
The correlation functions $K_{\ell\ell'}(\tau)=\lim_{t\to
\infty}\avg{\hatbf{I}_\ell(t+\tau)\hatbf{I}_{\ell'}(t)}$ can be
deduced from the master equation \eqref{eq:ME_mat}. In the matrix
notation they can be written as \cite{Hershfield93a,Korotkov94a}
\begin{multline}
\label{eq:K} K_{\ell\ell'}(\tau) = e^2\sum_{N,\set{n}}
\bra{N,\set{n}} \Bigg[ \Theta(+\tau)
\hatbf{I}_\ell\exp(-\whatbf{\Gamma}\tau)\hatbf{I}_{\ell'}
\\\mbox{}%
+ \Theta(-\tau)
\hatbf{I}_{\ell'}\exp(+\whatbf{\Gamma}\tau)\hatbf{I}_\ell
\\\mbox{}%
+ \delta(\tau)\delta_{\ell\ell'} \left(\whatbf{\Gamma}_\ell^+  +
\whatbf{\Gamma}_\ell^-\right) \Bigg] \ket{P(\infty)} \,,
\end{multline}
where $\Theta(x)$ is the unit step function.

To investigate the correlation effects, the noise power customarily
compared to the Poisson value $S_{Poisson}=2eI$. The Fano factor is
defined as the ratio of the actual noise power and the Poisson
value,
\begin{equation}
F\equiv\frac{S(0)}{2eI}. \label{eq:DefF}
\end{equation}

Since thermal noise ($S = 4k_B T G (V=0)$) is not particularly
interesting, we focus on the zero frequency shot noise, in the low
bias voltage regime where the Coulomb blockade governs the
electric transport; $T=0$ and $eV \lesssim 2E_C$.

We begin by considering analytically tractable cases with only a few
involved states, and then proceed to the full numerical results. The
finite frequency shot noise is briefly discussed in subsection
\ref{subsec:FFNoise}.

\subsection{Two-state model; $eV \approx \veps_p$ \label{subsec:SN_2state}}

The electron transport involving only two lowest energy states in
the quantum dot are well studied by many authors (see, for instance,
Ref. \onlinecite{Blanter00a}). Nevertheless, for later reference we
begin the discussion of shot noise with two-state process, which
provides a reasonable approximation for $eV\approx\veps_p$.  At
biases such that $eV\approx\veps_p$ and sufficiently low
temperatures, the two lowest states $\ket{N,n_1}=\ket{0,0}$ and
$\ket{1,0}$ dominate the transport process and the rate matrix is
given by
\begin{equation} \label{eq:G_2state}
\whatbf\Gamma=\left[\begin{array}{cc} \gamma^+ & -\gamma^-  \\
-\gamma^+ & \gamma^-
\end{array}\right]~,
\end{equation}
where the matrix elements are $\gamma^+ \equiv \Gamma_L(1,0\tol
0,0)$ and $\gamma^-\equiv \Gamma_R(0,0\tol 1,0)$.

With the current matrices defined by Eq. \eqref{eq:I_mat_L/R}
\begin{equation}
\hatbf I_L=\left[ \begin{array}{cc} 0 & 0  \\
\gamma^+ & 0 \end{array}\right]~,
~~\hatbf I_R=\left[ \begin{array}{cc} 0 & \gamma^-  \\
0 & 0 \end{array}\right]~, \n
\end{equation}
the noise power is obtained straightforwardly by Eqs.
\eqref{eq:Spower} and \eqref{eq:K} using the steady state
probability
\begin{equation} \label{eq:P_2state}
 \ket{P(\infty)}= \left[\begin{array}{c} P_{00} \\ P_{10}
  \end{array}\right] =\frac{1}{\gamma^+ +\gamma^-}
\left[\begin{array}{c} \gamma^- \\
\gamma^+ \end{array} \right].
\end{equation}
The Fano factor \eqref{eq:DefF} takes a simple form
\begin{equation} \label{eq:F_2state}
F_2 = P_{00}^2+P_{10}^2 =\frac{1+ (\gamma^-/\gamma^+)^2} {(1
+\gamma^-/\gamma^+)^2}
\end{equation}
where
\begin{equation}
\frac{\gamma^-}{\gamma^+} = \frac{1}{R}\left(\frac{eV/2-\delta
E_0}{eV/2+\delta E_0}\right)^\alpha \n
\end{equation}
and $\delta E_0$ is the shift of the bottom of the zero mode energy
induced by the gate voltage,
\begin{equation} \label{eq:delta_E0}
\delta E_0 =  (\delta N_G) \veps_p/g,~ \delta N_G = N_G-1/2.
\end{equation}
Note that Eq. \eqref{eq:F_2state} is valid for $|\delta E_0| \leq
eV/2$, otherwise $I=0$ and $S=0$ due to Coulomb blockade. We see from
Eq. \eqref{eq:F_2state} that the Fano factor is minimized for
$\gamma^-/\gamma^+=1$ and maximized for $(\gamma^-/\gamma^+)^{\pm
1}=0$, with the bounds $1/2 \leq F_2 < 1$. At the gate charge
$N_G=1/2$, it is determined only by the junction asymmetry parameter
$R$: $F_2=(1+R^2)/(1+R)^2$. Note that when only two states are
involved in the current carrying process (ground state to ground
state transitions), the Fano factor cannot exceed the Poisson value
$F=1$.

As a consequence of the power law dependence of the transition
rates on the transfer energy \eqref{eq:gamma_T0}, the Fano factor
is a function of the bias voltage, gate voltage, and the
interaction strength; it varies between the minimum and maximum values
\begin{equation} \label{eq:F_2state_b}
F_2 = \left\{\begin{array}{ll} \begin{displaystyle}
\frac{1}{2},\end{displaystyle} & ~ \mbox{ at }~
\begin{displaystyle} \delta E_0=\frac{1-R^{1/\alpha}}{1+R^{1/\alpha}}\frac{eV}{2}~
\end{displaystyle} \\
1,& ~\mbox{ at }~ \begin{displaystyle}\delta E_0 = \pm
\frac{eV}{2}~\end{displaystyle}
\end{array}\right.,
\end{equation}
({\em cf.} Eq. (5) in Ref.
\onlinecite{Braggio03a}).
\begin{figure}[htbp]
\includegraphics[width=8cm]{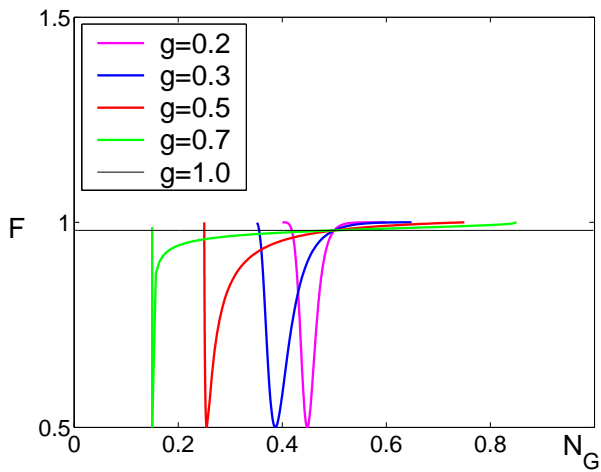}
\caption{(color online) Fano factor $F\equiv S(0)/2eI$ as a function of
  the gate
charge $N_G$ and LL interaction parameter $g$ (black line for
$g=1$, green $g=0.7$, red $g=0.5$, blue $g=0.3$, magenta $g=0.2$,
respectively) for $R=100$ (highly asymmetric junctions) at
$eV=\veps_p$ ($T=0$).} \label{fig:F_2state}
\end{figure}
Fig. \ref{fig:F_2state} depicts the Fano factor as a function of
gate charge $N_G$ and interaction parameter $g$ at the bias
$eV=\veps_p$ for strongly asymmetric junctions $R=100$ ($T=0$).
The gate charge $N_G$ corresponding to the minimum of the Fano factor
($F=1/2$) in the figure is
\begin{equation} \label{eq:NG_dip}
N_G^{\mathrm{dip}}=1/2+g(eV/2\veps_p)\frac{1-R^{1/\alpha}}{1+R^{1/\alpha}}.
\end{equation}
The Fano factor independently of the interaction strength crosses
$F_2=(1+R^2)/(1+R)^2$ at $N_G=1/2$, and it approaches maximum
$F_2=1$ at $N_G=1/2\pm g (eV/2\veps_p)$. i

\subsection{Three-state model; $eV\gtrsim 2\veps_p$ \label{subsec:SN_3state}}

The two-state model is applicable for bias voltages below
$eV_{\mathrm{th}} = 2(\veps_p -|\delta E_0|)$, since at least three
states can be involved in transport above this threshold voltage.
For electron transport involving three lowest energy states in the
quantum dot, $\ket{N,n_1}=\ket{0,0}, \ket{1,0},$ and $\ket{1,1}$
with $n_m=0$ for $m\geq 2$, the noise power can be calculated
exactly if the the contribution from the (backward) transitions
against the bias is negligible, as is typically the case at zero
temperature. In practice, however, the backward transitions are not
completely blocked for the bias above the threshold voltage of the
plasmon excitations, \emph{even at zero temperature}: once the bias
voltage reaches the threshold to initiate plasmon excitations, the
high energy plasmons in the QD above the Fermi energies of the leads
are also partially populated, opening the possibility of backward
transitions.

A qualitatively
new feature that can be studied in the three-state model as compared
to the two-state model is plasmon relaxation: the system with a
constant total charge may undergo transitions between different
plasmon configurations.

We will show in this subsection that the analytic solution of the
Fano factor of the three-state process yields an excellent agreement
with the low bias numerical results in the strong interaction
regime, while it shows small discrepancy in the weak interaction
regime (due to non-negligible contribution from the high energy
plasmons). We will also show that within the three state model the
Fano factor may exceed the Poisson value.


\subsubsection{Analytic results}

By allowing plasmon relaxation, the rate matrix involving three
lowest energy states $\ket{N,n_1}=\ket{0,0}, \ket{1,0},$ and
$\ket{1,1}$ is given by
\begin{equation} \label{eq:G_3state}
\whatbf\Gamma=\left[\begin{array}{ccc} \gamma_{0}^+ +\gamma_{1}^+
& -\gamma_0^- & -\gamma_{1}^- \\ -\gamma_0^+ & \gamma_0^- &
-\gamma_p  \\ -\gamma_1^+ & 0 & \gamma_1^- +\gamma_p
\end{array}\right]~,
\end{equation}
with the matrix elements $\gamma_{i}^+ \equiv \Gamma_L(1,i\tol
0,0), \gamma_i^-\equiv \Gamma_R(0,0\tol 1,i),~i=0,1$ and
$\gamma_p$ introduced in Eq. \eqref{eq:Gp}. Current matrices
defined by Eq. \eqref{eq:I_mat_L/R} are
\[[\hatbf I_L]_{\{2,1\}}=\gamma_0^+,[\hatbf I_L]_{\{3,1\}}=\gamma_1^+,
[\hatbf I_R]_{\{1,2\}}=\gamma_0^-, [\hatbf
I_R]_{\{1,3\}}=\gamma_1^- ~,\]
with $[\hatbf I_\ell]_{\{i,j\}}=0$ for other set of $i,j=1,2,3$,
where $\ell=L,R$.
The noise power is obtained straightforwardly by Eqs.
\eqref{eq:Spower} and \eqref{eq:K} using the steady state
probability
\begin{equation}
 \ket{P(\infty)}= \left[\begin{array}{c} P_{00} \\ P_{10} \\ P_{11}
  \end{array}\right] =\frac{1}{Z}
\left[\begin{array}{c} \gamma_0^-(\gamma_1^- +\gamma_p) \\
\gamma_0^+(\gamma_1^- + \gamma_p) +\gamma_1^+\gamma_p \\
\gamma_1^+\gamma_0^- \end{array} \right],
\end{equation}
with normalization constant $Z=\gamma_0^+\gamma_1^- +
\gamma_1^+\gamma_0^- + \gamma_0^-\gamma_1^- + (\gamma_0^+
+\gamma_0^- +\gamma_1^+)\gamma_p$.
Using the average current $I=e(P_{10}\gamma_0^- +
P_{11}\gamma_1^-)$, the Fano factor $F=S(0)/2eI$ is given by
\begin{multline} \label{eq:F_3state}
F_3 = P_{00}^2+P_{10}^2+P_{11}^2  +
2P_{11}\frac{1}{Z}\frac{\gamma_0^+}{\gamma_0^-} \\ \times
\bigg[(\gamma_0^-)^2 + (\gamma_1^-)^2-\gamma_0^-\gamma_1^-
+\frac{\gamma_0^+ + \gamma_1^+}{\gamma_0^+}
\gamma_1^-\gamma_p\bigg]~.
\end{multline}
Compared to the Fano factor \eqref{eq:F_2state} in the two-state
process, complication arises already in the three-state process
due to the last term in Eq. \eqref{eq:F_3state} which results from
the coupling of $P_{11}$ and the rates which cannot be expressed
by the components of the probability vector.

In order to have $\ket{N,n_1}=\ket{0,0}, \ket{1,0},$ and
$\ket{1,1}$ as the relevant states, we assume $\gamma_0^+
> \gamma_0^-$ or more explicitly $N_G\geq N_G^{\mathrm{dip}}$ which
is introduced in Eq. \eqref{eq:NG_dip} . In the opposite situation
($\gamma_0^+ < \gamma_0^-$), the relevant states are
$\ket{N,n_1}=\ket{0,0}, \ket{0,1},$ and $\ket{1,0}$, and above
description is still valid with the exchange of electron number
$N=0 \leftrightarrow 1$ and the corresponding notations
$\gamma_i^+ \leftrightarrow \gamma_i^-,~i=0,1$.

\begin{figure}[htbp]
\includegraphics[width=8cm]{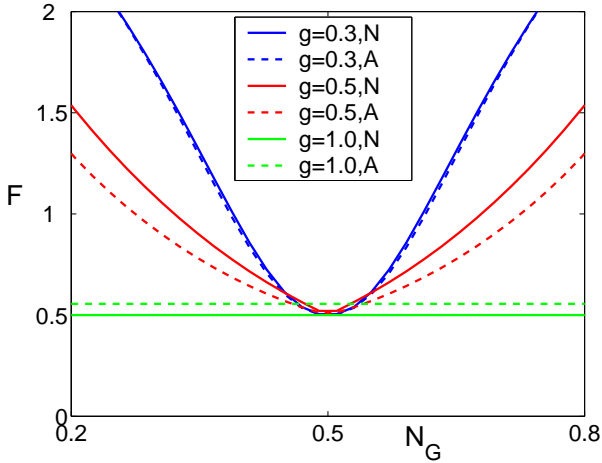}
\caption{(color online) Fano factor $F\equiv S(0)/2eI$ as a function of
  the gate
charge $N_G$ for symmetric junctions ($R=1$) at voltage $eV=2\veps_p$
($T=0$), with no plasmon relaxation. Numerical results (solid
lines) vs. analytic results with three states, Eq.
\eqref{eq:F_3state} (dashed lines) for $g=0.3$ (blue), $0.5$
(red), and $1.0$ (green).} \label{fig:FvsNGsym}
\end{figure}

To see the implications of Eq. \eqref{eq:F_3state}, we plot the
Fano factor in Fig. \ref{fig:FvsNGsym}, with respect to the gate
charge $N_G$ for symmetric junctions at $eV=2\veps_p$ ($T=0$),
with no plasmon relaxation ($\gamma_p=0)$.

Two main features are seen in Fig. \ref{fig:FvsNGsym}. Firstly,
the shot noise is enhanced over the Poisson limit
($F=1$) in the strong interaction regime, $g\lesssim 0.5$, for a
range of parameters with gate charges near (but not including) $N_G=1/2$. As
discussed above, in the low bias regime $eV<2(\veps_p-|\delta E_0|)$ at
zero temperature, no plasmons are excited and the electric charges
are transported via only the two-state process following the Fano
factor \eqref{eq:F_2state} which results in the sub-Poissonian
shot noise ($1/2\leq F\leq 1$). Once the bias reaches the
threshold $eV_{\mathrm{th}}= 2(\veps_p-|\delta E_0|)$, it
initiates plasmon excitations which enhance the shot noise
over the Poisson limit. This feature is discussed
in more detail below.

Secondly, in the weak interaction regime ($g\gtrsim 0.5$) a small
discrepancy between the analytic result \eqref{eq:F_3state} (dashed
line) and the numerical result (solid line) is found. It results
from the partially populated states of the high energy plasmons over
the bias due to non-vanishing transition rates. On the other hand,
a simple three-state approximation shows excellent agreement in the
strong interaction regime ($g=0.3$ in the figure), indicating
negligible contribution of the high energy plasmons ($E_D(N,\{n\})
> eV/2$) to the charge transport mechanism. This is due to the power law
suppression of the transition rates \eqref{eq:gamma_T0} as a
function of the transfer energy \eqref{eq:W_L/R}.

\subsubsection{Limiting cases}

To verify the role of non-equilibrium plasmons as the cause of the
shot noise enhancement, we consider two limiting cases of Eq.
\eqref{eq:F_3state}: $\gamma_p=0$ and $\gamma_p \gg \gamma_i^\pm$.

In the limit of no plasmon relaxation ($\gamma_p=0$), the Fano
factor \eqref{eq:F_3state} of the three-state process is
simplified as
\begin{multline} \label{eq:F_3state_limit1}
F_3^{(0)} = 1-2[(1-P_{10})P_{10}+(1-P_{11})P_{11}]
 \\ +2P_{10}P_{11}\frac{(\gamma_0^-)^2 +
 (\gamma_1^-)^2}{\gamma_0^-\gamma_1^-},
\end{multline}
with steady-state probability
\begin{equation}
 \ket{P(\infty)}= \left[\begin{array}{c} P_{00} \\ P_{10} \\ P_{11}
  \end{array}\right] =\frac{1}{Z}
\left[\begin{array}{c} \gamma_0^-\gamma_1^- \\
\gamma_0^+\gamma_1^-  \\ \gamma_1^+\gamma_0^- \end{array} \right],
\end{equation}
where the new normalization constant is $Z=\gamma_0^+\gamma_1^- +
\gamma_1^+\gamma_0^- + \gamma_0^-\gamma_1^-$.

The three-state approximation is most accurate in the low bias
regime $2(\veps_p-|\delta E_0|)\lesssim eV \ll 2E_C$ ($\delta E_0$
is defined in Eq. \eqref{eq:delta_E0}) and for gate voltages away
from $N_G=1/2$, i.e., for $1/2<N_G\lesssim 1$ (or $0\lesssim
N_G<1/2$ with the exchange of indices regarding particle number $N=0
\leftrightarrow 1$). In this regime, $\gamma_0^+$ and/or
$\gamma_1^-$ dominate over the other rates, ($\gamma_0^+,
\gamma_1^-) \gg (\gamma_{0}^-,\gamma_{1}^+$), which results in $1
\gtrsim P_{10} \gg (P_{0}, P_{11}) \approx 0$. The Fano factor
\eqref{eq:F_3state_limit1} now reduces to
\begin{equation} \label{eq:F_3state_limit1b}
F_3^{(0)}\approx 1+2P_{10}P_{11}\frac{(\gamma_0^-)^2 +
 (\gamma_1^-)^2}{\gamma_0^-\gamma_1^-} \approx 1
 +2\frac{\gamma_1^+}{\gamma_0^+}.
\end{equation}
Notice that while Eq. \eqref{eq:F_3state_limit1} is an exact
solution for the three-state process with no plasmon relaxation,
Eq. \eqref{eq:F_3state_limit1b} is a good approximation only
sufficiently far
from $N_G= 1/2$. In this range, Eq.
\eqref{eq:F_3state_limit1b} explicitly shows that \emph{the
opening of new charge transport channels accompanied by the
plasmon excitations causes the enhancement of the shot noise (over
the Poisson limit)}.

In the limit of fast plasmon relaxation, on the other hand,
$\gamma_p \gg \gamma_i^\pm$ and effectively no plasmon is excited,
\begin{equation}
 \ket{P(\infty)}\approx \left[\begin{array}{c} P_{00} \\ P_{10} \\ P_{11}
  \end{array}\right] =\frac{1}{\gamma_0^+ +\gamma_1^+ + \gamma_0^-}
\left[\begin{array}{c} \gamma_0^- \\
\gamma_0^+ +\gamma_1^+  \\ 0 \end{array} \right].
\end{equation}
Consequently, the Fano factor \eqref{eq:F_3state} is given by
\begin{equation} \label{eq:F_3state_limit2}
F_3^{(\infty)}\approx \frac{1}{2}
+2\left(P_{00}-\frac{1}{2}\right)^2 \in [\frac{1}{2},1].
\end{equation}
The maximum Fano factor $F=1$ is reached if one of the rates
$\gamma_0^+$ or $\gamma_0^-$ dominates, while the minimum value
$F=1/2$ requires that $\gamma_0^- = \gamma_0^+ + \gamma_1^+$, i.e.,
that the total tunneling-in and tunneling-out rates are equal;more
explicitly,
\begin{equation}
\frac{1}{R}\left(1-\frac{2\delta E_0}{eV/2 + \delta
E_0}\right)^\alpha
-\frac{1}{g}\left(1-\frac{\varepsilon_p}{eV/2+\delta
E_0}\right)^\alpha = 1.
\end{equation}

\begin{figure}[htbp]
\includegraphics[width=8cm]{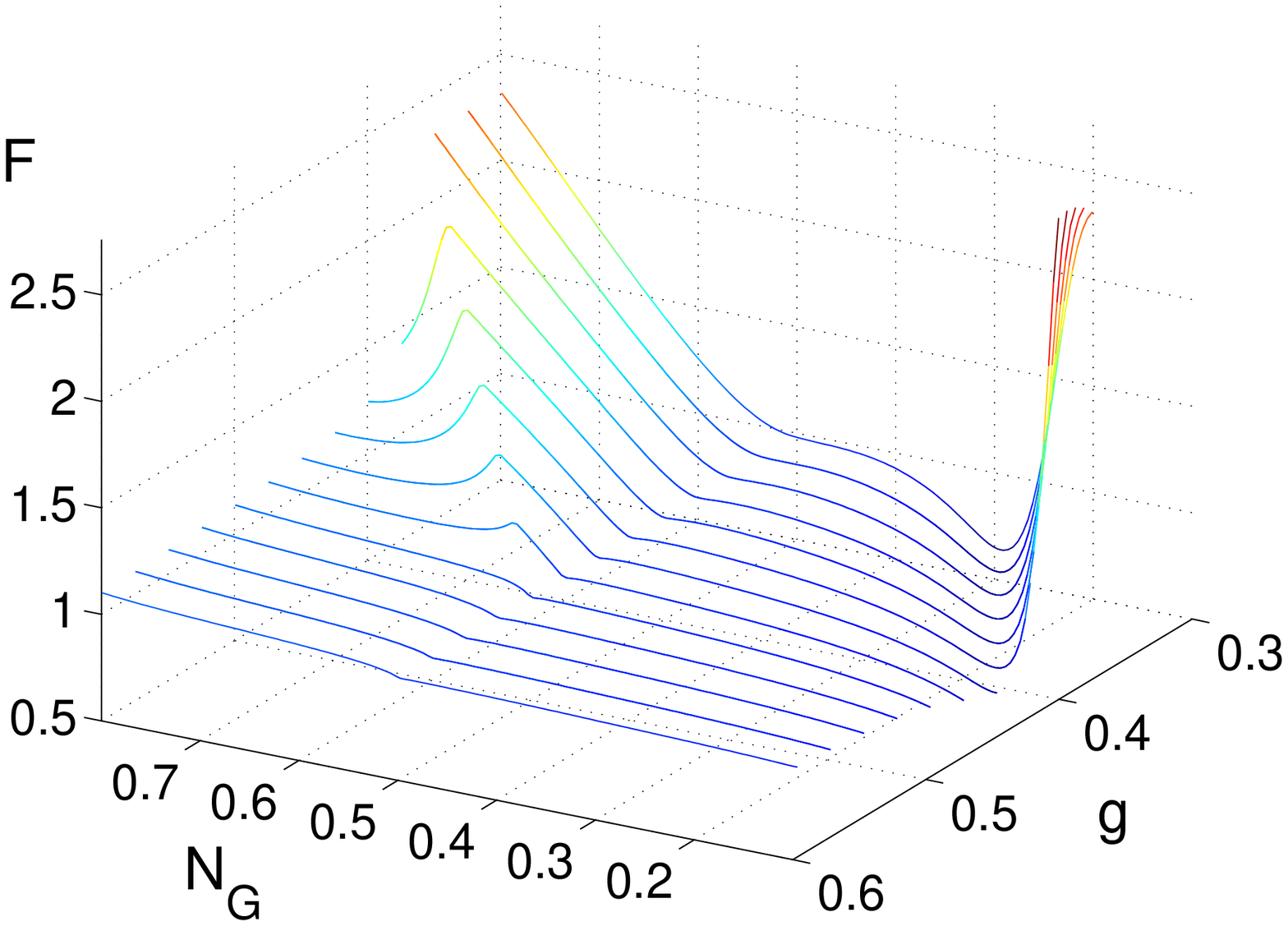}\\ (a) \\
\includegraphics[width=8cm]{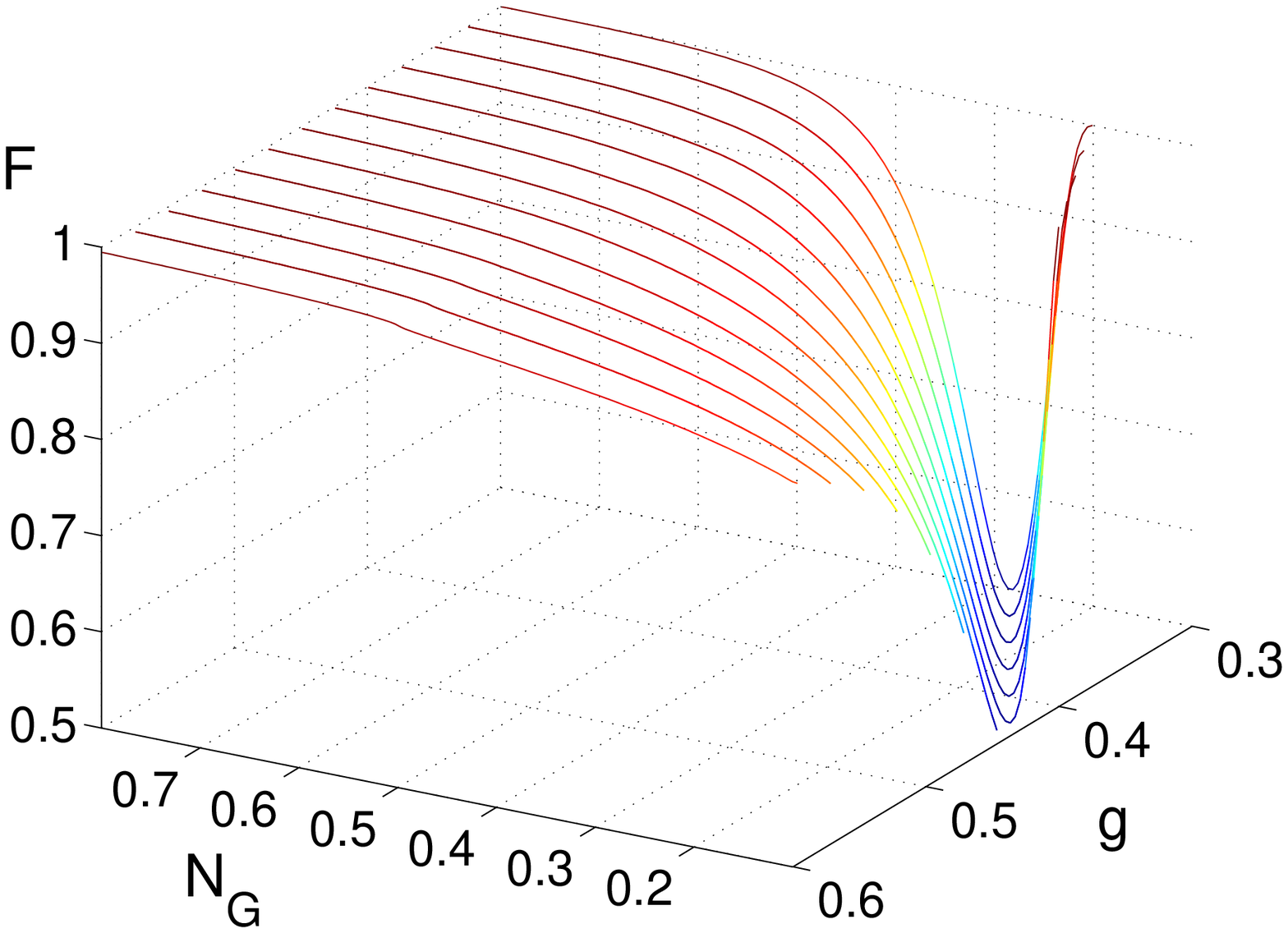}\\ (b)
\caption{(color online) Fano factor $F\equiv S(0)/2eI$ as a function of
  the gate
charge $N_G$ and LL interaction parameter $g$ for $R=100$ (highly
asymmetric junctions) at $eV=2\veps_p$ ($T=0$), (a) with no
plasmon relaxation ($\gamma_p=0$) and (b) with fast plasmon
relaxation ($\gamma_p=10^4$). } \label{fig:FvsNGvsg_asym}
\end{figure}

\subsection{Numerical results}

The limiting cases of no plasmon relaxation ($\gamma_p=0$) and a
fast plasmon relaxation ($\gamma_p=10^4$) are summarized in Fig.
\ref{fig:FvsNGvsg_asym}(a) and (b), respectively. In the figure
the Fano factor is plotted as a function of gate voltage and
interaction parameter $g$ in the strong interaction regime
$0.6\leq g\leq 0.3$ at $eV=2\veps_p$ for $R=100$ (strongly
asymmetric junctions).

As shown in Fig. \ref{fig:FvsNGvsg_asym}(a), the Fano factor is
enhanced above the Poisson limit ($F=1$) for a range of gate charges
away from $N_G = 1/2$, especially in the strong interaction regime,
as expected from the three-state model. The shot noise enhancement
is lost in the presence of a fast plasmon relaxation process, in
agreement with analytic arguments, as seen in Fig.
\ref{fig:FvsNGvsg_asym}(b) when $F$ is bounded by $1/2 \leq F \leq
1$. Hence, slowly relaxing plasmon excitations enhance shot noise,
and this enhancement is most pronounced in the strongly interacting
regime.

In the limit of fast plasmon relaxation $F$ exhibits a minimum value
$F=1/2$ at positions consistent with predictions of the three-state
model: the voltage polarity and ratio of tunneling matrix elements
at the two junctions is such that total tunneling-in and
tunneling-out rates are roughly equal for small values of $N_G$. If
plasmon relaxation is slow, $F$ still has minima at approximately
same values of $N_G$ but the minimal value of the Fano factor is
considerably larger due to the presence of several transport
channels.

\subsection{Interplay between charge fluctuations and plasmon excitations near $eV=2E_C$
\label{subsec:ChargeFluctuation}}

So far, we have investigated the role of non-equilibrium plasmons as
the cause of the shot noise enhancement and focused on a voltage
range when only two charge states are significantly involved in
transport. The question naturally follows what is the consequence of
the charge fluctuations. Do they enhance shot noise, too?

To answer this question, we first consider a toy model in which
the plasmon excitations are absent during the single-charge
transport. In the three-N-state regime where the relevant states
are $\ket{N}=\ket{-1},\ket{0}$ and $\ket{1}$ with no plasmon
excitations at all. At zero temperature, the rate matrix in this
regime is given by
\begin{equation} \label{eq:G_3Nstate}
\whatbf\Gamma=\left[\begin{array}{ccc} \Gamma_{-1}^+  &
-\Gamma_0^-
& 0 \\ -\Gamma_{-1}^+ & \Gamma_0^- + \Gamma_0^+ & -\Gamma_1^-  \\
0 & -\Gamma_0^+ & \Gamma_1^-
\end{array}\right]~,
\end{equation}
where the matrix elements are $\Gamma_i^{+}=\Gamma_L(i+ 1,\{0\} \tol
i,\{0\}), \Gamma_i^{-}=\Gamma_R(i- 1,\{0\} \tol i,\{0\}),
~i=-1,0,1$. Repeating the procedure in subsection
\ref{subsec:SN_3state}, we arrive at a Fano factor that has a
similar form as Eq. \eqref{eq:F_3state_limit1},
\begin{multline} \label{eq:F_3Nstate}
F_{3N} = 1-2[(1-P_{-1})P_{-1}+(1-P_{1})P_{1}]
 \\ +2P_{-1}P_{1}\left(\frac{\Gamma_{-1}^+}{\Gamma_{1}^-}
 + \frac{\Gamma_{1}^-}{\Gamma_{-1}^+} \right),
\end{multline}
with the steady-state probability vector
\begin{equation}
 \ket{P(\infty)}= \left[\begin{array}{c} P_{-1} \\ P_{0} \\ P_{1}
  \end{array}\right] =\frac{1}{Z}
\left[\begin{array}{c} \Gamma_0^-\Gamma_1^- \\
\Gamma_{-1}^+\Gamma_1^- \\
\Gamma_{-1}^+\Gamma_0^+ \end{array} \right]
\end{equation}
where $Z=\Gamma_0^-\Gamma_1^- +\Gamma_{-1}^+\Gamma_1^-
+\Gamma_{-1}^+\Gamma_0^+$.

Despite the formal similarity of Eq. \eqref{eq:F_3Nstate} with Eq.
\eqref{eq:F_3state_limit1}, its implication is quite different. In
terms of the transition rates, $F_{3N}$ reads
\begin{multline} \label{eq:F_3Nstate_b}
F_{3N} = 1-\frac{2}{Z^2}\Big[\Gamma_1^-\Gamma_0^- (\Gamma_{-1}^+
\Gamma_0^+ +(\Gamma_{-1}^+ -\Gamma_0^+)\Gamma_1^-) \\
+ \Gamma_{-1}^+\Gamma_0^+ (\Gamma_{0}^- \Gamma_1^- +(\Gamma_{1}^-
-\Gamma_0^-)\Gamma_1^-) \Big].
\end{multline}
Since  $\Gamma_{-1}^+ > \Gamma_0^+$ and $\Gamma_{1}^-
>\Gamma_0^-$ in the three-N-state regime, the Fano factor $F_{3N}$ is
sub-Poissonian, i.e., $F_{3N} < 1$,  consistent with the
conventional equilibrium descriptions
\cite{Hershfield93a,Braggio03a}.

We conclude that while plasmon excitations may enhance the shot
noise over the Poisson limit, charge fluctuations, in contrast, do
not alter the sub-Poissonian nature of the Fano factor in the low
energy regime. This qualitative difference is due to the fact that
certain transition rates between different charge states vanish
identically (in the absence of co-tunneling): it is impossible for
the system to move directly from a state with $N=-1$ to $N=1$ or
vice versa.

\begin{figure}[htbp]
\includegraphics[width=8cm]{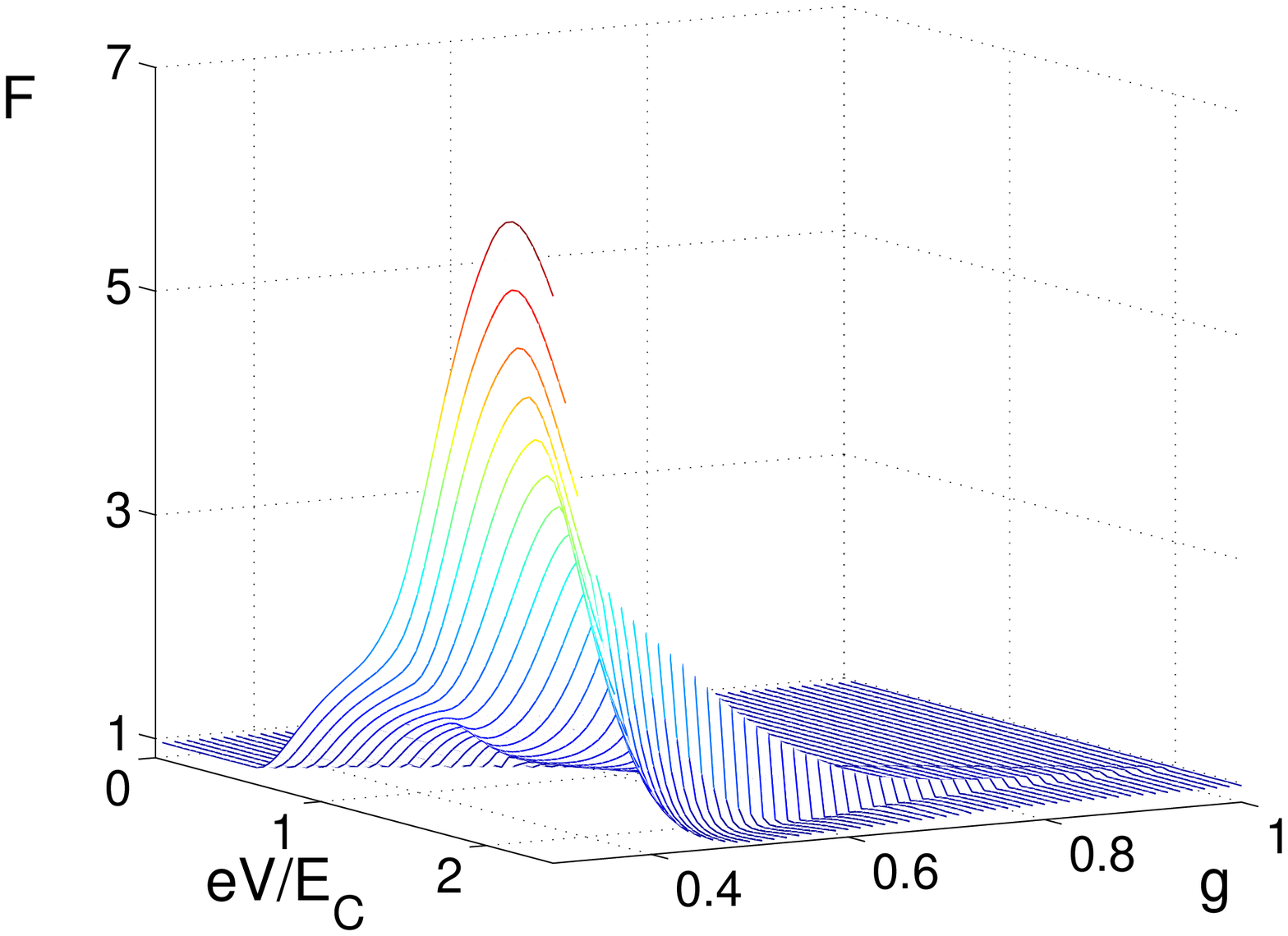}\\ (a) \\
\includegraphics[width=8cm]{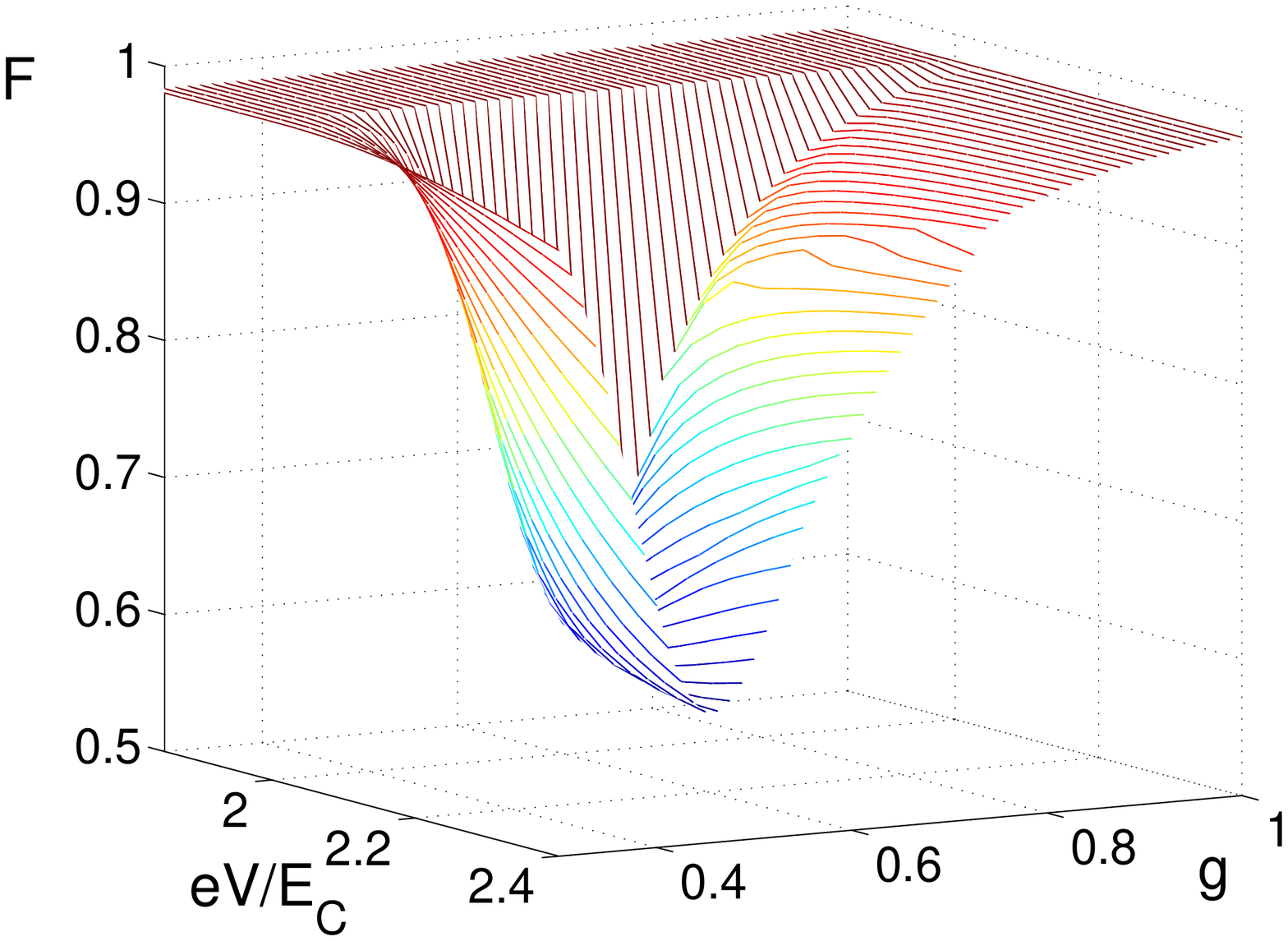}\\ (b)
\caption{(color online) Fano factor $F\equiv S(0)/2eI$ as a function of the bias
$eV$ and LL interaction parameter $g$ for $R=100$ (highly
asymmetric junctions) at $N_G=1/2$ ($T=0$), (a) with no plasmon
relaxation ($\gamma_p=0$) and (b) with fast plasmon relaxation
($\gamma_p=10^4$). } \label{fig:FeVg}
\end{figure}

Therefore, we expect that for bias voltages near $eV =2E_C >
2\veps_p$, when both plasmon excitations and charge fluctuations
are relevant, the Fano factor will exhibit complicated
non-monotonic behavior. Exact solution is not tractable in this
regime since it involves too many states. Instead, we calculate
the shot noise numerically, with results depicted in Fig.
\ref{fig:FeVg}, where the zero temperature Fano factor is shown as
a function of the bias $eV$ and LL interaction parameter $g$ for
$R=100$ at $N_G=1/2$, (a) with no plasmon relaxation
($\gamma_p=0$) and (b) with fast plasmon relaxation
($\gamma_p=10^4$).

In the bias regime up to the charging energy $eV\leq 2E_C$, the Fano
factor increases monotonically due to non-equilibrium plasmons. On
the other hand, the charge fluctuations contribute at $eV\geq 2E_C$
which tends to suppress the Fano factor. As a consequence of this
competition, the Fano factor reaches its peak at $eV = 2E_C$ and is
followed by a steep decrease as shown in Fig. \ref{fig:FeVg}(a).
Note the significant enhancement of the Fano factor in the strong
interaction regime, which is due in part to the power law dependence
of the transition rates with exponent $\alpha=(1/g-1)$ as discussed
earlier, and in part to more plasmon states being involved for
smaller $g$ since $E_C = \veps_p/g$. The latter reason also accounts
for the fact that the Fano factor begins to rise at a lower apparent
bias for smaller $g$: the bias voltage is normalized by $E_C$ so
that plasmon excitation is possible for lower values of $eV/E_C$ for
stronger interactions.

In the case of fast plasmon relaxation, the rich structure of the
Fano factor due to non-equilibrium plasmons is absent as shown in
Fig. \ref{fig:FeVg}(b), in agreement with the discussions in
previous subsections. The only remaining structure is a sharp dip
around $eV= 2E_C$ that can be attributed to the charge fluctuations
at $eV\geq 2E_C$. Not only is the minimum value of the Fano factor a
function of the interaction strength but also the bias voltage at
which it occurs depends on $g$. The minimum Fano factor occurs at
higher bias voltage, and the dip tends to be deeper with increasing
interaction strength. Note for $g\lesssim 0.5$, the Fano factor did
not reach its minimum still at largest voltages plotted
($eV/(2E_C)=1.2$).

The Fano factor at very low voltages $eV<2\veps_p$ for $N_G=1/2$
is $F^{(0)}=(1+R^2)/(1+R)^2$ (see Eq. \eqref{eq:F_2state}),
regardless of the plasmon relaxation mechanism. As shown in Fig.
\ref{fig:FeVg}(b), the Fano factor is bounded above by this value
in the case of fast plasmon relaxation. Notice that in the case of
no plasmon relaxation (\ref{fig:FeVg}(a)), the dips in the Fano
factor at $eV>2E_C$ reach below $F^{(0)}$. See Fig. 1(a) in Ref.
\onlinecite{KimJU04a} for more detail.

Since both the minimum value of $F$ and the voltage at which it
occurs are determined by a competition between charge fluctuations
and plasmonic excitations, they cannot be accurately predicted by
any of the simple analytic models discussed above.

\subsection{Finite frequency noise \label{subsec:FFNoise}}

\begin{figure}[htbp]
\includegraphics[width=8cm]{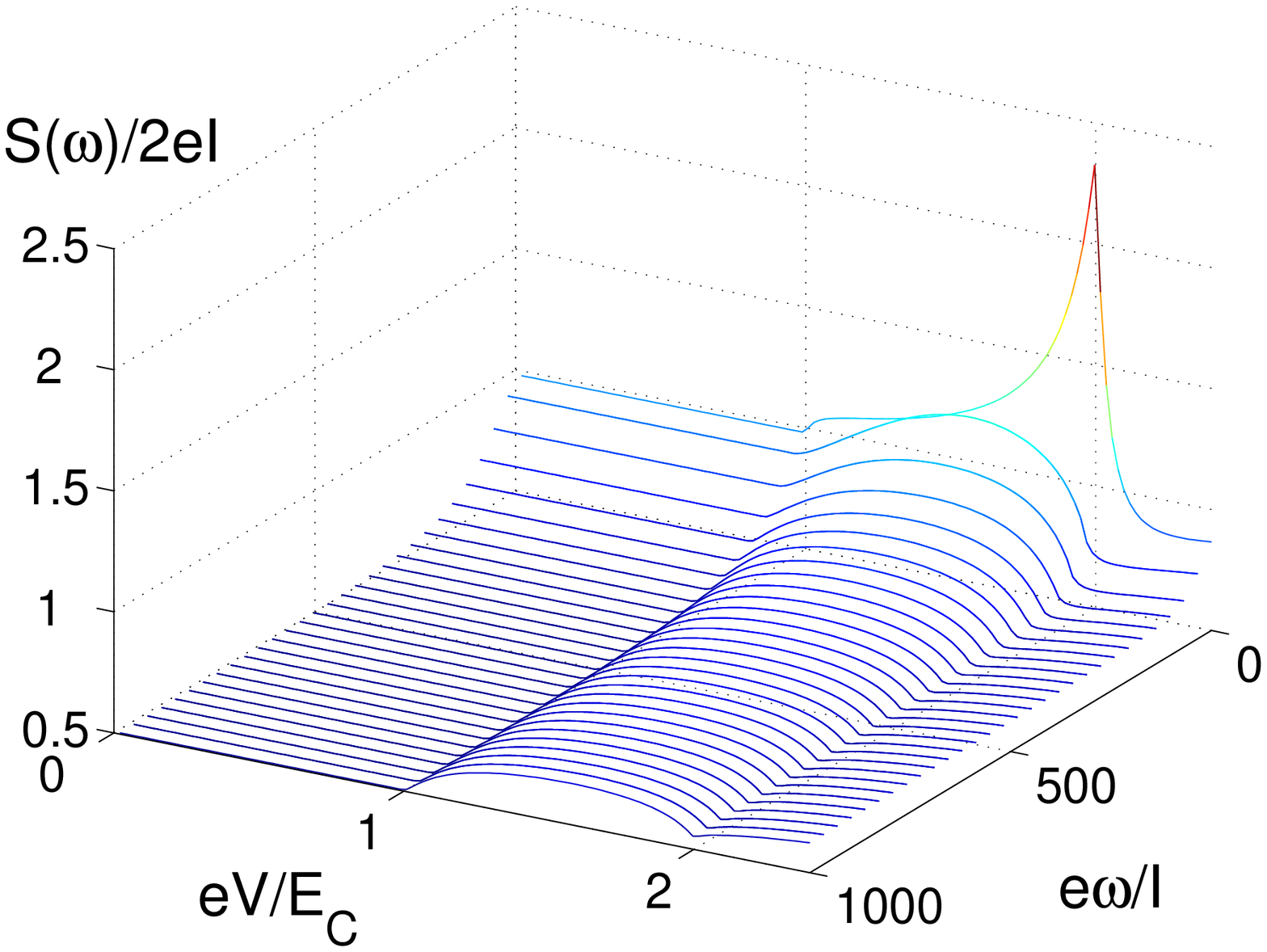}\\ (a) \\
\includegraphics[width=8cm]{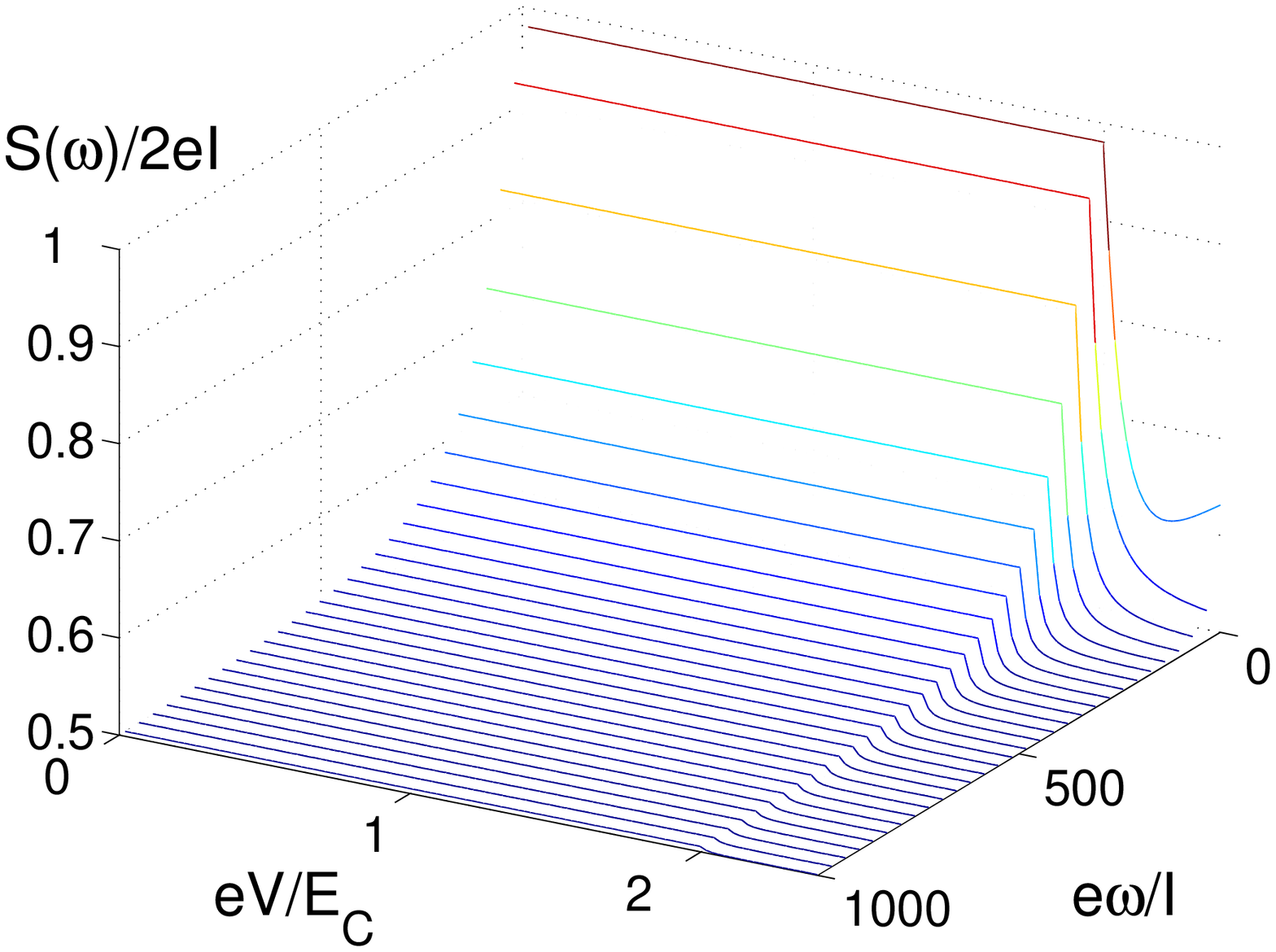}\\ (b)
\caption{(color online) Finite frequency shot noise $S(\w)/2eI$ as a function of
the bias $eV/E_C$ and frequency $e\w/I$ for $R=100$ and $g=0.5$,
at $N_G=1/2$ ($T=0$), (a) with no plasmon relaxation
($\gamma_p=0$) and (b) with fast plasmon relaxation
($\gamma_p=10^4$). } \label{fig:FWeV}
\end{figure}

In the high frequency limit, $\omega \rightarrow \infty$, the
correlation effects are lost in the noise power \eqref{eq:Spower}
except the $\delta(\tau)$-term in Eq. \eqref{eq:K} that
reflects the Pauli exclusion \cite{Korotkov92a}, and the
asymptotic value of the noise spectrum reduces to
\begin{equation} \label{eq:SN_infty}
S(\infty) = 2e\left(\frac{C_R^2A_L}{(C_L+C_R)^2}+
\frac{C_L^2A_R}{(C_L+C_R)^2}\right) =\frac{e}{2}(A_L+A_R),
\end{equation}
where
\begin{equation}
\label{Noise-llj-long::eq:A} A_{\ell}=e \sum_{N,\{n\}} \langle
N,\{n\}|\whatbf{\Gamma}^+_{\ell}
+\whatbf{\Gamma}^-_{\ell}|P(\infty)\rangle
\end{equation}
is the total tunneling rate across the junction $\ell$ without
regard to direction.

The decay of the current-current correlations at finite frequencies is depicted
in Fig. \ref{fig:FWeV}, where the shot noise power $S(\w)/2eI$ is shown
as a function of the bias and the frequency for a strong
interaction ($g=0.5$) and the strongly asymmetric tunnel barriers
($R=100)$ (a) with no plasmon relaxation and (b) with fast plasmon
relaxation ($\gamma_p=10^4$).

In the regime of the \textit{elastic} process, in which the charge
transport does not involve excitations (plasmons)
in the dot, the backward tunneling against the bias is blocked at
zero temperature. In this regime, the rate density is equal to the
average current, $A_L=A_R=I$, and the high frequency noise power
asymptotically converges to its minimum value $S(\infty)=eI$. This
is clearly seen in the regime of the low bias $eV<2\veps_p(=E_C)$
in Fig. \ref{fig:FWeV}(a) and (b).

In the presence of fast plasmon relaxation, still no backward
tunneling is possible and the shot noise reaches its minimum value
$S(\infty)=eI$ at high frequency limit (see Fig. \ref{fig:FWeV}(b)).
However, the non-equilibrium plasmons excited above the Fermi
energies of the leads invoke non-vanishing backward tunneling and
the high frequency shot noise remains above $eI$, as shown in the
bias regime of $eV > 2\veps_p$ in Fig. \ref{fig:FWeV}(a).

Another identifiable feature in Fig. \ref{fig:FWeV}(a) is the rapid
decrease of the shot noise power as a function of $\omega$ at those
voltages when the Fano factor has a maximum. The maxima occur at
voltages when additional states become significantly populated, and
the characteristic low frequency reflects the slow transition rates
to states near their energy thresholds.

\section{Full counting statistics \label{sec:FCS}}

Since shot noise that is a current-current correlation is more
informative than the average current, we expect more information
with higher-order currents or charge correlations. The method of
counting statistics, which was introduced to mesoscopic physics by
\citet{Levitov93a} followed by \citet{Muzykantskii94a} and
\citet{LeeH95a}, shows that all orders of charge correlation
functions can be obtained as a function related to the probability
distribution of transported electrons for a given time interval.
This powerful approach is known as full counting statistics (FCS).
The first experimental study of the third cumulant of the voltage
fluctuations in a tunnel junction was carried out by
\citet{Reulet03a}. The experiment indicates that the high
cumulants are more sensitive to the coupling of the system to the
electromagnetic environment. See also \citet{Levitov04a} and
\citet{Beenakker03b} for the theoretical discussions on the third
cumulant in a tunnel-barrier.

\citet{deJong96a} showed that the low-energy physics of the FCS
calculations both with quantum mechanical and semiclassical
approaches are identical in the DB structures, as in the explicit
shot noise calculations, except for a short initial time scale.

We will now carry out an FCS analysis of transport through a
double barrier quantum wire system. The analysis will provide
a more complete characterization of the transport properties
of the system than either average current or shot noise, and
shed further light on the role of non-equilibrium {\em vs.}
equilibrium plasmon distribution in this structure.

Let $P(M,\tau)$ be the probability that $M$ electrons have
tunneled across the right junction to the right lead during the
time $\tau$. We note that
\begin{multline*} P(M,\tau) = \lim_{t\to\infty}
\sum_{N,\set{n}}\sum_{M_0,N_0,\set{n_0}}
\\ P(M_0+M,N,\set{n},t+\tau;M_0,N_0,\set{n_0},t),
\end{multline*}
where $P(M_0+M,N,\set{n},t+\tau;M_0,N_0,\set{n_0},t)$ is called
joint probability since it is the probability {\em that}, up to time
$t$, $M_0$ electrons have passed across the right junction and
$N_0$ electrons are confined in the QD with $\set{n_0}$ plasmon
excitations, {\em and that} $M_0+M$ electrons have passed R--junction with
$(N,\{n\})$ excitations in the QD up to time $t+\tau$. The master
equation for the joint probability can easily be constructed from
Eq. \eqref{eq:ME_mat} by noting that $M\to M\pm 1$ as $N\to N\mp
1$ via only R--junction hopping.

 To obtain $P(M,\tau)$, it is convenient to define the characteristic
 function conjugate to the joint probability as
\begin{multline*}
 g(\theta,N,\set{n},\tau) = \lim_{t\to\infty} \sum_M
e^{-i\theta M} \sum_{M_0,N_0,\set{n_0}} \\
P(M_0+M,N,\set{n},t+\tau;M_0,N_0,\set{n_0},t) .
\end{multline*}

The characteristic function satisfies the master equation
\begin{equation}
\label{eq:ME3} \frac{\partial}{\partial\tau}\ket{g(\theta,\tau)} =
-\whatbf\Gamma(\theta)\ket{g(\theta,\tau)}
\end{equation}
with the initial condition
\begin{math}
\ket{g(\theta,\tau=0)} = \ket{P(\infty)}
\end{math}.
The $\theta$-dependent $\whatbf\Gamma$ in Eq.~\eqref{eq:ME3} is
related to the previously defined transition rate matrices through
\begin{equation}
\label{eq:G21} \whatbf\Gamma(\theta) = \widehat\Gamma_p +
\left(\whatbf\Gamma_L^0 - \whatbf\Gamma_L^+ -
\whatbf\Gamma_L^-\right) + \left(\whatbf\Gamma_R^0 -
\whatbf\Gamma_R^+ e^{+i\theta}
  - \whatbf\Gamma_R^- e^{-i\theta}\right) \,.
\end{equation}
The characteristic function $G(\theta,\tau)$ conjugate to
$P(M,\tau)$ is now given by
\begin{math}
G(\theta,\tau) = \sum_{N,\set{n}}
\braket{N,\set{n}|g(\theta,\tau)}
\end{math},
or
\begin{equation}
\label{eq:G_char} G(\theta,\tau) = \sum_{N,\set{n}}
\bra{N,\set{n}}\exp\left[-\whatbf\Gamma(\theta)\tau\right]
\ket{P(\infty)} \,.
\end{equation}
Finally, the probability $P(M,\tau)$ is obtained by
\begin{equation}
\label{eq:PM} P(M,\tau) = \int_{0}^{2\pi}\frac{d\theta}{2\pi}\;
e^{+i\theta M}G(\theta,\tau) = \oint \frac{dz}{2\pi
i}\frac{G(z,\tau)}{z^{2M+1}}
\end{equation}
with $z=e^{-i\theta/2}$, where the contour runs counterclockwise
along the unit circle and we have used the symmetry property
$G(z,\tau)=G(-z,\tau)$ for the second equality.

Taylor expansion of the logarithm of the characteristic function
in $i\theta$ defines the \emph{cumulants} or irreducible
correlators $\k_k(\tau)$:
\begin{equation}
\ln G(\theta,\tau) = \sum_{k=1}^\infty \frac{(i \theta)^k}{k!}
\k_k(\tau) \: . \label{eq:Fcs_cumulants}
\end{equation}
The cumulants have a direct polynomial relation with the moments
$\overline{n^k(\tau)}\equiv \sum_n n^k P(n,\tau)$. The first two
cumulants are the mean and the variance, and the third cumulant
characterizes the asymmetry (or skewness) of the $P(M,\tau)$
distribution and is given by
\begin{equation} \label{eq:Fcs_k3}
\k_3(\tau)= \overline{\d n(\tau)^3
}=\overline{(n(\tau)-\overline{n(\tau)})^3}.
\end{equation}

In this section, we investigate FCS mainly in the context of the
probability $P(M,\tau)$ that M electrons have passed through the
right junction during the time $\tau$. Since the average current
and the shot noise are proportional to the average number of the
tunneling electrons $\avg{M}$ and the width of the distribution of
$P(M,\tau)$, respectively, we focus on the new aspects that are
not covered by the study of the average current or shot noise.

In order to get FCS in general cases we integrate the master
equation (\ref{eq:ME3}) numerically (see subsection
\ref{subsec:FCS_numerical}). In the low-bias regime, however, some
analytic argument can be made. We will show through the following
subsections that for symmetric junctions in the low-bias
regime ($2\veps _p<eV<2E_C$), and irrespective of the junction
symmetry in the very low bias regime ($eV < 2 \veps_p$),
$P(M,\tau)$ is given by the residue at $z=0$ alone,
\begin{equation}
\label{eq:PM_G_char} P(M,\tau) =
\frac{1}{(2M)!}\left.\frac{d^{2M}}{dz^{2M}}\right|_{z=0} G(z,\tau)
\, .
\end{equation}
Through this section we assume that the gate charge is $N_G=1/2$,
unless stated explicitly not so.

We will now follow the outline of the previous section and start
by considering two analytically tractable cases before proceeding
with the full numerical results.

\subsection{Two-state process; $eV \approx \veps_p$ \label{subsec:FCS_2state}}

For the very low bias $eV< 2 \veps_p$ at zero temperature, no
plasmons are excited and electrons are carried by transitions
between two states $(N,n)=(0,0) \leftrightarrow (1,0)$. In this
simplest case, the rate matrix $\whatbf\Gamma(\theta)$ in Eq.
\eqref{eq:ME3} is determined by only two participating transition
rates $\gamma^+ = \Gamma_L(1,0\tol 0,0)$ and $\gamma^- =
\Gamma_R(0,0\tol 1,0)$,
\begin{equation} \label{eq:G_theta_2state}
\whatbf\Gamma(\theta)=\left[\begin{array}{cc} \Gamma_0+\delta &
-(\Gamma_0-\delta)e^{-i\theta}\\
-(\Gamma_0+\delta) & \Gamma_0-\delta
\end{array}\right]
\end{equation}
with $\Gamma_0\equiv(\gamma^+ +\gamma^-)/2,~\delta\equiv (\gamma^+
-\gamma^-)/2$.

Substituting the steady-state probability Eq. \eqref{eq:P_2state}
and the transition rate matrix \eqref{eq:G_theta_2state} to Eq.
\eqref{eq:G_char}, one finds
\begin{multline} \label{eq:G_char2}
G_2(z,\tau) = \frac{e^{-\Gamma_0\tau}}{4} \frac{1}{f_2(z)}  \bigg[
(1+ f_2(z))^2 e^{+\Gamma_0\tau f_2(z)} \\
- (1- f_2(z))^2 e^{-\Gamma_0\tau f_2(z)} \bigg],
\end{multline}
where $f_2(z)=(\Gamma_\delta/\Gamma_0)\sqrt{z^2+\Delta^2}$, with
$\Gamma_\delta =
\sqrt{\Gamma_0^2-\delta^2}=\sqrt{\gamma^+\gamma^-}$ and $\Delta^2
= \delta^2/\Gamma_\delta^2$.

Now, it is straightforward to calculate the cumulants. In the long
time limit $\tau\gg (\gamma^{\pm})^{-1}$, for instance, in terms
of the average current $I_2= e\gamma^+\gamma^-/(\gamma^+
+\gamma^-)$ and the Fano factor $F_2$ in \eqref{eq:F_2state}, the
three lowest cumulants are given by
\begin{eqnarray} \label{eq:Fcs_2s_k}
&&\kappa_1(\tau) = I_2\tau, \quad
\kappa_2(\tau) = e I_2 F_2 \tau, \n\\
&&\kappa_3(\tau) = e^2 I_2 (3(F_2-1/2)^2+1/4)\tau,
\end{eqnarray}
where the electron charge ($-e$) is revived. These are in agreement
with the phase-coherent quantum-mechanical results \cite{deJong96a}.
It is convenient to discuss the asymmetry (skewness) by the ratio
$A\equiv\kappa_3/e^2 \kappa_1$ ($\tau\to\infty$), noticing the Fano
factor $F=\lim_{\tau\to \infty} \kappa_2(\tau)/e\kappa_1(\tau) $.
The factor $A_2= 3(F_2-1/2)^2+1/4$ is positive definite (positive
skewness) and bounded by $1/4\leq A_2\leq 1$. It is interesting to
notice that $A_2$ is a monotonic function of $F_2$ and has its
minimum $A_2=1/4$ for the minimum $F_2=1/2$ and maximum $A_2=1$ for
the maximum $F_2=1$. Notice $A=1$ for a Poissonian and $A=0$ for a
Gaussian distribution. Therefore, the dependence of $A_2$ on the
gate charge $N_G$ is similar to that of the Fano factor $F_2$ (see
Fig. \ref{fig:F_2state}) with dips at $N_G^{\mathrm{dip}}$ in Eq.
\eqref{eq:NG_dip}. Together with Eq. \eqref{eq:F_2state_b}, it
implies that the effective shot noise and the asymmetry of the
probability distribution per unit charge transfer have their
respective minimum values at the gate charge $N_G^{\mathrm{dip}}$
which depends on the tunnel-junction asymmetry and the interaction
strength of the leads.

\begin{figure}[htbp]
\includegraphics[height=6.5cm]{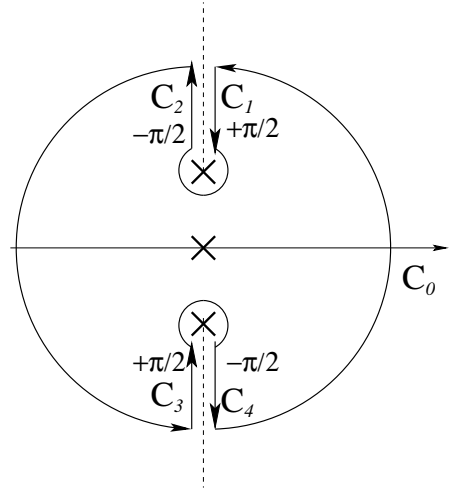}
\caption{Contour of Eq. \eqref{eq:PM}. The arguments of
$f_2(z)=(\Gamma_\delta/\Gamma_0)\sqrt{z^2+\Delta^2}~$ are
$\pi/2,-\pi/2,\pi/2,-\pi/2$ along the branch cuts $C_1,C_2,C_3$,
and $C_4$, respectively.} \label{fig:FCStour}
\end{figure}
The integral in Eq. \eqref{eq:PM} is along  the contour depicted
in Fig. \ref{fig:FCStour}. Notice that the contributions from the
part along the branch cuts are zero and we are left with the
multiple poles at $z=0$. By residue theorem, the two-state
probability $P_2(M,\tau)$ is given by Eq. \eqref{eq:PM_G_char}.

The exact expression of $P_2(M,\tau)$ is cumbersome. For symmetric
tunneling barriers  with $N_G=1/2$
($\delta=0,~\gamma^+=\gamma^-=\Gamma_0$), however, Eq.
\eqref{eq:G_char2} reduces to
\begin{multline} \label{eq:G_char2s}
G_2^{(s)}(z,\tau) = \frac{e^{-\Gamma_0\tau}}{4z}\Big( (1+z)^2
e^{+\Gamma_0\tau z}- (1-z)^2 e^{-\Gamma_0\tau z}\Big).
\end{multline}
Accordingly, $P^{(s)}_2(M,\tau)$ is concisely given by
\begin{multline} \label{eq:PM_2s}
P^{(s)}_2(M,\tau) = e^{-\Gamma_0\tau}\bigg[\frac{1}{2}
\frac{(\Gamma_0\tau)^{2M-1}}{(2M-1)!}+ \frac{(\Gamma_0\tau)^{2M}}{(2M)!} \\
+\frac{1}{2}\frac{(\Gamma_0\tau)^{2M+1}}{(2M+1)!} \bigg], \mbox{
for }~ M\geq 1,
\end{multline}
with $P^{(s)}_2(0,\tau) = e^{-\Gamma_0\tau}( 1+\Gamma_0\tau/2)$, in
agreement with Eq. (24) of Ref. \onlinecite{deJong96a}. While this
distribution resembles a sum of three Poisson distributions, it is
not exactly Poissonian.

For a highly asymmetric junctions $R\gg 1$ $(\gamma^+ \gg
\gamma^-)$, the first term in Eq. \eqref{eq:G_char2} dominates the
dynamics of $G_2(z,\tau)$ and its derivatives, and the
characteristic function is approximated by
\begin{equation} \label{eq:G_char2as}
G_2^{(a)}(z,\tau) \approx \exp\left(-\tau \Gamma_0+\tau \sqrt{
(\gamma^+\gamma^- )z^2+\delta^2} \right).
\end{equation}
Now, the solution of $P^{(a)}_2(M,\tau)$ is calculated by this
equation and Eq. \eqref{eq:PM_G_char}. The leading order
approximation in $\gamma^-/\gamma^+$ leads to the Poisson
distribution,
\begin{equation}\label{eq:PM_2a}
P^{(a)}_2(M,\tau)\approx \frac{(\gamma^- \tau)^M}{M!} e^{-\gamma^-
\tau}.
\end{equation}
For a single tunneling-barrier, the charges are transported by the
Poisson process ($F=1$) \cite{Blanter00a}. Therefore, we recover the
Poisson distribution in the limit of strongly asymmetric junctions
and in the regime of the two-state process, in which electrons see
effectively single tunnel-barrier.

For the intermediate barrier asymmetry, therefore, the probability
$P_2(M,\tau)$ of a two-state process is given by a distribution
between Eq. \eqref{eq:PM_2s} (for symmetric--junctions) and the
Poissonian \eqref{eq:PM_2a} (for the most asymmetric--junctions).

\subsection{Four-state process; $2\veps_p \lesssim eV \lesssim 2E_C$
 \label{subsec:FCS_4state}}

Since we focus the FCS analysis on the case $N_G = 1/2$, the
next simplest case to study is a four-state-model rather than the
three-state-model discussed in the connection of the shot noise in
the previous section.

In the bias regime where the transport is governed by Coulomb
blockade ($eV < 2E_C$) and yet the plasmons play an important role
($eV \geq 2 \veps_p$), it is a fairly good approximation to include
only the four states with $N=0,1$  and $n_1=0,1$
($n_m=0$ for $m\geq 2$). For general asymmetric cases, the rate
matrix $\whatbf\Gamma(\theta)$ in Eq. \eqref{eq:ME3} is determined
by ten participating transition rates (four rates from each
junction, and two relaxation rates). The resulting eigenvalues of
$\ket{g(\theta,\tau)}$ are the solutions of a quartic equation,
which is in general very laborious to solve analytically.

For symmetric junctions ($R = 1$) with no plasmon relaxation,
however, the rate matrix is simplified to
\begin{equation} \label{eq:G_theta_4state}
\whatbf\Gamma(\theta)=\left[\begin{array}{cccc}
\gamma_{00}+\gamma_{10} & 0 & -\gamma_{00} z^2 & -\gamma_{01}z^2
\\ 0 & \gamma_{01}+\gamma_{11} & -\gamma_{10} z^2 &
-\gamma_{11}z^2\\ -\gamma_{00} & -\gamma_{01} &
\gamma_{00}+\gamma_{10} & 0 \\ -\gamma_{10} & -\gamma_{11} & 0 &
\gamma_{01}+\gamma_{11}
\end{array}\right]
\end{equation}
with the matrix elements $\gamma_{ij}\equiv \Gamma_L(1,i\tol
0,j)=\Gamma_R(0,i\tol 1,j)$. The steady-state probability is then
given by
\begin{equation} \label{eq:P4infty}
 \ket{ P_4^{(s)}(\infty)}=\ket{ g(z,\tau=0)} =
 \frac{1}{2(\gamma_{01}+\gamma_{10})}
 \left[ \begin{array}{l} \gamma_{01} \\ \gamma_{10} \\ \gamma_{01}
 \\ \gamma_{10}\end{array}\right]
\end{equation}
Solving Eq. \eqref{eq:G_char} with this probability and the rate
matrix \eqref{eq:G_theta_4state} is laborious but straightforward
and one finds
\begin{multline} \label{eq:G_char4sym}
G(z,\tau) = e^{-(\gamma_{00}+\gamma_{10}+ \gamma_{01}+\gamma_{11})
\tau/2}  \\ \times \bigg[\Big\{G_I(z,\tau)
\frac{(1+z)^2}{8z}\Big\} +\Big\{z\to -z\Big\}\bigg],
\end{multline}
with
\begin{multline}
G_I(z,\tau) = \bigg\{e^{(\gamma_{00}z+\gamma_{11}z +
f_4(z))\tau/2}\left(1+\frac{A(z)}{f_4(z)}\right) \\
\times
\left(1-\frac{A(z)-f_4(z)}{2(\gamma_{01}+\gamma_{10})z}\right)\bigg\}
+\bigg\{f_4(z)\to -f_4(z)\bigg\}, \label{eq:G_char4sym2}
\end{multline}
where $A(z)$ and $f_4(z)$ are given by
\begin{eqnarray}
A(z) &=&\gamma_{00}+\gamma_{10}-\gamma_{01}- \gamma_{11} -
(\gamma_{00}-\gamma_{11}-2\gamma_{01}) z \n\\
f_4(z) &=& \sqrt{(\gamma_{00}-\gamma_{11})^2-4
\gamma_{10}\gamma_{01}} \sqrt{(z-a)^2+b^2}
\label{eq:BranchPoint4state}
\end{eqnarray}
with dimensionless parameters $a,b$ given by
\begin{eqnarray}
a &=& \frac{(\gamma_{00}-\gamma_{11})
(\gamma_{00}+\gamma_{10}-\gamma_{01}-\gamma_{11})}
{(\gamma_{00}-\gamma_{11})^2-4 \gamma_{10}\gamma_{01}}, \n\\
b &=& \frac{\sqrt{4\gamma_{10}\gamma_{01}}
(\gamma_{00}+\gamma_{10}-\gamma_{01}-\gamma_{11}))}
{(\gamma_{00}-\gamma_{11})^2-4 \gamma_{10}\gamma_{01}}. \n
\end{eqnarray}
Integral of $G_I(z,\tau)$ along the contour $|z|=1$ contains two
branch points at $z_c=a\pm ib$, however, the integral along the
branch cuts cancel out due to the symmetry under $[f_4(z)\to
-f_4(z)]$. Therefore, the contribution from the branch cuts due to
$G_I(z,\tau)$ and $G_I(-z,\tau)$ is zero to the probability
$P_4^{(s)}(M,\tau)$, and it is given by the residues only at
$z=0$, i.e. by Eq. \eqref{eq:PM_G_char}. The explicit expression
of $P_4^{(s)}(M,\tau)$ is cumbersome.



The probability distribution $P_2(M,\tau)$ for the two-state
process deviates from Eq. \eqref{eq:PM_2s} as a function of the
asymmetry parameter $R$ and reaches Poissonian in the case of
strongly asymmetric--junctions. In a similar manner,
$P_4^{(s)}(M,\tau)$ deviates from $P_2^{(s)}(M,\tau)$ as a
function of ratio of the transition rates $\gamma_{ij}$.

\subsection{Numerical results \label{subsec:FCS_numerical}}

It is worth mentioning that for strongly asymmetric junctions
$P(M,\tau)$ is Poissonian in the very low bias regime
($eV<2\veps_p$), as seen in Eq. \eqref{eq:PM_2a}. It exhibits a
crossover at $eV=2\veps_p$: $P(M,\tau)$ deviates from Poisson
distribution for $2\veps_p<eV<2E_C$ while it is Poissonian for
$eV<2\veps_p$ (at $T=0$), as shown by the shot noise calculation.

\subsubsection{Voltage dependence \label{subsubsec:ECS_eV}}

\begin{figure}[htbp]
\includegraphics[width=8cm]{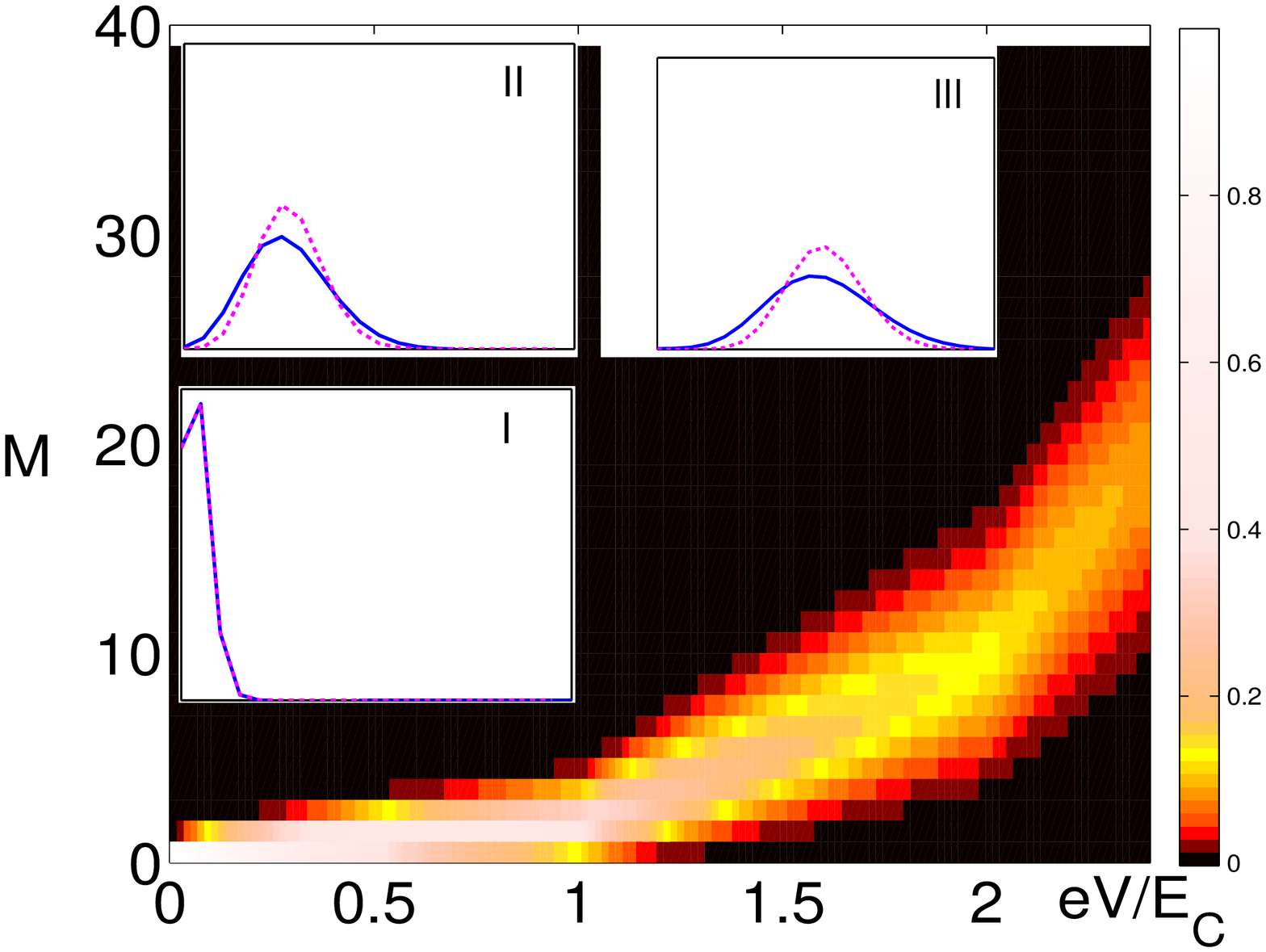}\\
(a) \\
\includegraphics[width=8cm]{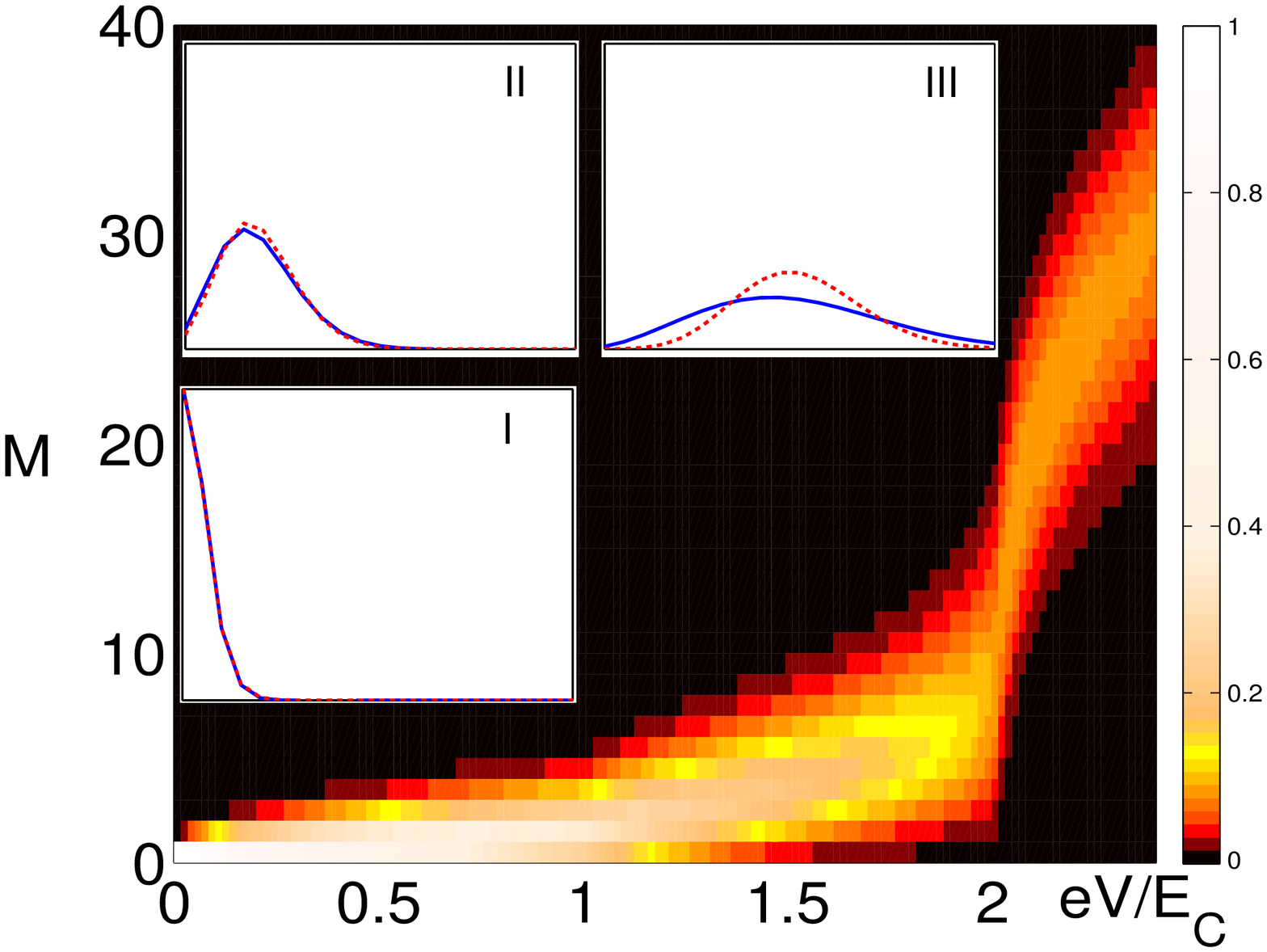}\\
(b)
 \caption{(color online) Probability $P(M,\tau)$ with no plasmon relaxation
($\gamma_p=0$) during the time $\tau = 10/I_c$, where $I_c$ is the
particle current at $eV=2E_C$ with no plasmon relaxation
($\gamma_p = 0$); (a) for symmetric junctions ($R=1$) and (b) for
a highly asymmetric junctions ($R=100$). Here $g=0.5$, $N_G=1/2$,
and $T = 0$. Inset shows cross-sectional image of $P(M,\tau)$
(blue solid line) as a function of $M$ and the reference
distribution function (a) Eq. \eqref{eq:PM_2s} (magenta dashed
line) and (b) the Poisson distribution Eq. \eqref{eq:PM_2a} (red
dashed), at \rm{(I)} $eV=0.5 E_C$, \rm{(II)} $eV=1.5 E_C$, and
\rm{(III)} $eV=2 E_C$. } \label{fig:Fcs3D_PMeV}
\end{figure}

The analytic results presented above are useful in interpreting the
numerical results in Fig. \ref{fig:Fcs3D_PMeV}, where probability
$P(M,\tau)$ for (a) symmetric junctions ($R=1$) and (b) $R=100$, in
the case of LL parameter $g=0.5$ with no plasmon relaxation
($\gamma_p=0$), is shown as a function of $eV$ and $M$ that is the
number of transported electrons to the right lead during $\tau$ such
that during this time $\avg{M}=I_{c}\tau=10$ electrons have passed
to the right lead at $eV=2E_C$.

The peak position of the distribution of
$P(M,\tau)$ is roughly linearly proportional to the average particle flow
$\avg{M}$, and the width is proportional to the shot noise
but in a nonlinear manner. In a rough estimate, therefore, the
ratio of the peak width to the peak position is proportional to
the Fano factor. Two features are shown in the figure. First, the
average particle flow (the peak position) runs with different
slope when the bias voltage crosses new energy levels, i.e. at
$eV=2\veps_p$ and $eV=2E_C$, that is consistent with the $I-V$
study (compare Fig. \ref{fig:Fcs3D_PMeV}(b) with Fig.
\ref{fig:IV_R100}). Notice $E_C=\veps_p/g=2\veps_p$ for $g=0.5$.
Second, the width of the distribution increases with increasing
voltage, with different characteristics categorized by
$eV=2\veps_p$ and $eV=2E_C$. Especially in the bias regime
$eV>2E_C$ in which the charge fluctuations participate to the
charge transport, for the highly asymmetric case, the peak runs
very fast while its width does not show noticeable increase. It
causes the dramatic peak structure in the Fano factor around
$eV=2E_C$ as discussed in section \ref{sec:noise}.


The deviation of the distribution of probability $P(M,\tau)$ due to
the non-equilibrium plasmons from its low voltage (equilibrium)
counterpart is shown in the insets. Notice in the low bias regime
$eV< 2\veps_p$, it follows Eq. \eqref{eq:PM_2s} for the symmetric
case (\ref{fig:Fcs3D_PMeV}(a), inset \rm{I}), and the Poissonian
distribution \eqref{eq:PM_2a} for the highly asymmetric case
(\ref{fig:Fcs3D_PMeV}(b), inset \rm{I}). The deviation is already
noticeable at $eV=3 \veps_p=1.5 E_C$ for $R=1$ (inset \rm{II} in
\ref{fig:Fcs3D_PMeV}(a)), while it deviates strongly around
$eV=2E_C$ for $R=100$ (inset \rm{III} in \ref{fig:Fcs3D_PMeV}(a)).

\subsubsection{Interaction strength dependence \label{subsubsec:ECS_g}}

\begin{figure}[htbp]
\includegraphics[width=8cm]{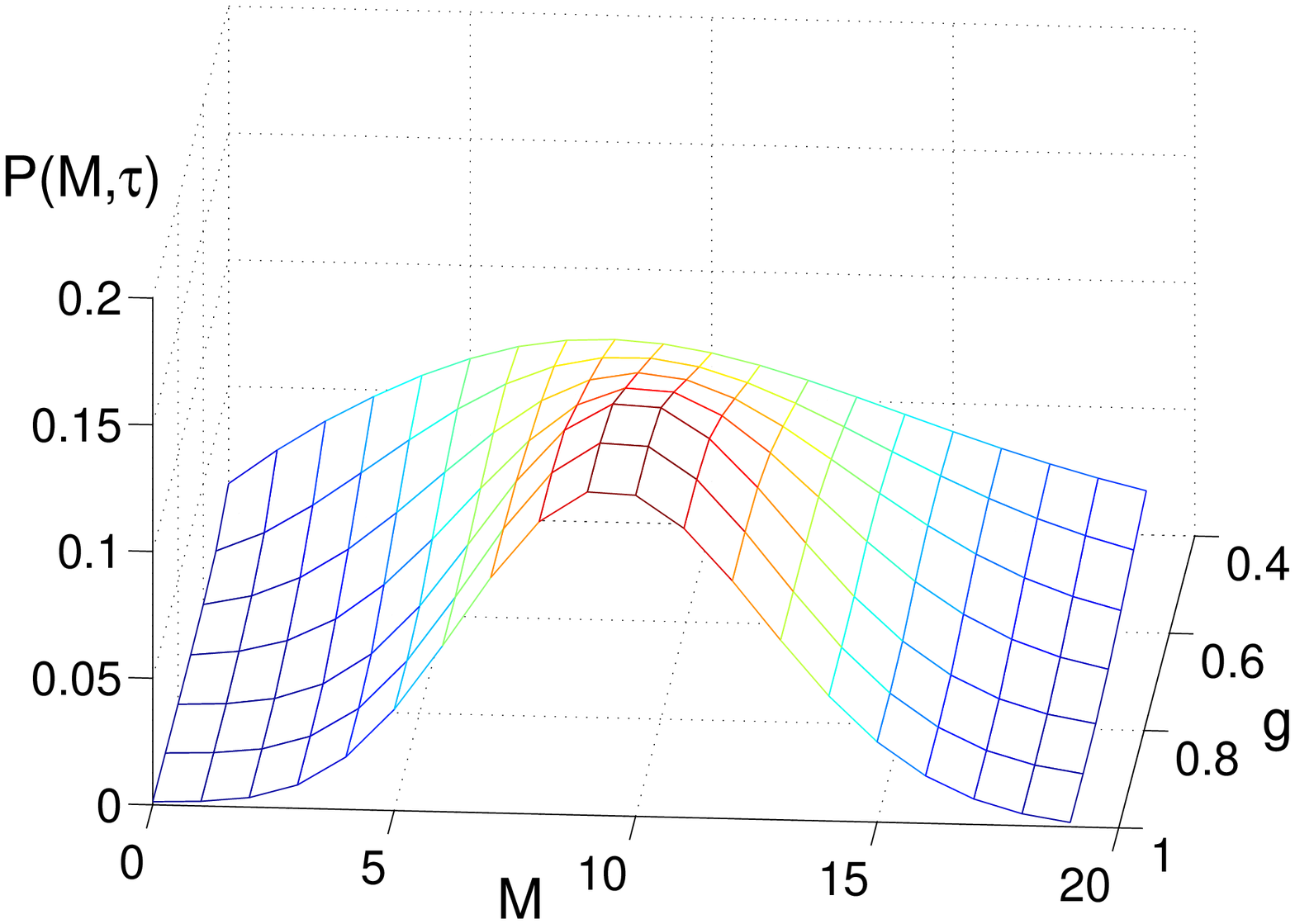}
\\ (a) \\
\includegraphics[width=8cm]{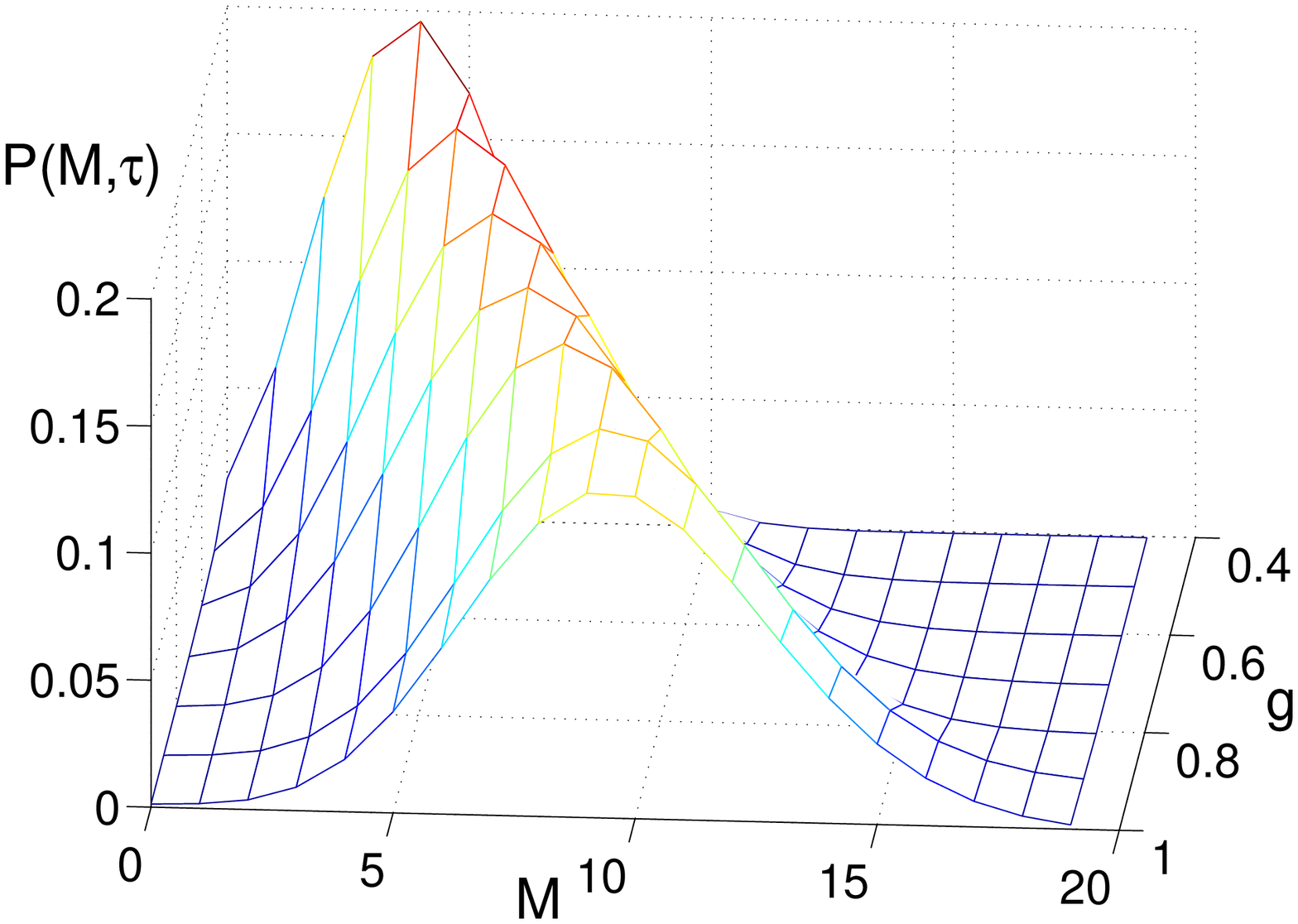}
\\ (b)
\caption{(color online) The probability $P(M,\tau)$ that M electrons have
passed through the right junction during the time $\tau = 10/I_0$,
where $I_0$ is the particle current with no plasmon relaxation
($\gamma_p = 0$); (a) with no plasmon relaxation ($\gamma_p=0$)
and (b) with fast plasmon relaxation ($\gamma_p=10^4$). Here $eV =
2E_C, R = 100, N_G=1/2$ and $T = 0$.}
\label{fig:Fcs3D_PMg}
\end{figure}

We have concluded in section \ref{sec:noise} that shot noise shows
most dramatic behavior around $eV=2E_C$ due to interplay between the
non-equilibrium plasmons and the charge fluctuations. To see its
consequence in FCS, we plot in Fig. \ref{fig:Fcs3D_PMg} the
probability $P(M,\tau)$ as a function of the particle number $M$ and
the interaction parameter $g$ for $\tau = 10e/I(eV=2E_C,\gamma_p=0)$
(a) with no plasmon relaxation and (b) with fast plasmon relaxation.

The main message of Fig. \ref{fig:Fcs3D_PMg}(a) is that the shot
noise enhancement, i.e. the broadening of the distribution curve, is
significant in the strong interaction regime with gradual increase
with decreasing $g$. Fast plasmon relaxation consequently suppresses
the average current dramatically as shown in Fig.
\ref{fig:Fcs3D_PMg}(b) implying the Fano factor enhancement is lost.
Effectively, the probability distribution of $P(M,\tau)$ for
different interaction parameters maps on each other almost
identically if the time duration is chosen such that $\avg{M}$
equals for all $g$.


\section{Conclusions \label{sec:conclusions}}

We have studied different transport properties of a Luttinger-liquid
single-electron transistor including average current, shot noise,
and full counting statistics,
within the conventional sequential
tunneling approach.

At finite bias voltages, the occupation probabilities of the
many-body states on the central segment is found to follow a highly
non-equilibrium distribution.
The energy is transferred between the leads and the quantum dot by
the tunneling electrons, and the electronic identity is dispersed
into the plasmonic collective excitations after the tunneling
event.
In the case of nearly symmetric barriers, the distribution of the
occupation probabilities of the non-equilibrium plasmons shows
impressive contrast depending on the interaction strength: In the
weakly interacting regime, it is a complicated function of the
many-body occupation configuration, while in the strongly
interacting regime, the occupation probabilities are determined
almost entirely by the state energies and the bias voltage, and
follow a universal distribution resembling Gibbs (equilibrium)
distribution. This feature in the strong interaction regime fades
out with the increasing asymmetry of the tunnel--barriers.

We have studied the consequences of these non-equilibrium plasmons
on the average current, shot noise, and counting statistics. Most
importantly, we find that the average current is increased, shot
noise is enhanced beyond the Poisson limit, and full counting
statistics deviates strongly from the Poisson distribution. These
non-equilibrium effects are pronounced especially in the strong
interaction regime, i.e. $g\lesssim 0.5$. The overall transport
properties are determined by a balance between phenomena associated
with non-equilibrium plasmon distribution that tend to increase
noise, and charge fluctuations that tend to decrease noise. The
result of this competition is, for instance, a non-monotonic voltage
dependence of the Fano factor.

At the lowest voltages when charge can be transported through the
system, the plasmon excitations are suppressed, and the Fano
factor is determined by charge fluctuations. Charge fluctuations
are maximized when the tunneling-in and tunneling-out rates are
equal, which for symmetric junctions occurs at gate charge
$N_G=1/2$. At these gate charges the Fano factor acquires its
lowest value which at low voltages is given by a half of the
Poisson value, known as $1/2$ suppression, as only two states are
involved in the transport, at somewhat larger voltages increases
beyond the Poisson limit as plasmon excitations are allowed, and
at even higher voltages exhibits a local minimum when additional
charge states are important. If the non-equilibrium plasmon
effects are suppressed e.g. by fast plasmon relaxation, only the
charge fluctuation effects survive, and the Fano factor is reduced
below its low-voltage value. The non-equilibrium plasmon effects
are also suppressed in the non-interacting limit.

\begin{acknowledgments}
This work has been supported by the Swedish Foundation for
Strategic Research through the CARAMEL consortium, STINT, the SKORE-A
program, the eSSC at Postech, and the SK-Fund.

\end{acknowledgments}

\appendix
\section{The transition amplitudes in the quantum dot \label{app:OndotT}}

In this appendix, we derive the transition amplitudes,
Eqs. \eqref{eq:Laguerre} and \eqref{eq:phi_nu}.
As shown in Eq. \eqref{eq:gamma_D}, the zero-mode overlap of the QD
transition amplitude is unity or zero. Therefore, we focus on the
overlap of the plasmon states. It is enough to consider $$ |\langle
\{n^\prime\}|\Psi^\dagger_{D}(X_\ell)|\{n\}\rangle|^2,~~X_\ell=X_L,X_R
$$ due to the symmetry between matrix elements of tunneling-in and
tunneling-out transitions, as in Eq. \eqref{eq:Laguerre},
\begin{widetext}
\begin{eqnarray}\label{eq:app_DotT}
|\langle \{n^\prime\}|\Psi^\dagger(X_\ell)|\{n\}\rangle|^2
&=&~\sum_{r=\pm} \Big[ |\langle
\{n^\prime\}|\psi^\dagger_{r}(X_\ell)|\{n\}\rangle|^2  +\langle
\{n^\prime\}|\psi^\dagger_{r}(X_\ell)|\{n\}\rangle \langle \{n\}
|\psi_{-r}(X_\ell)| \{n^\prime\}\rangle \Big]  \n\\
&=&2 |\langle \{n^\prime\}|\psi^\dagger_{r}(X_\ell)|\{n\}\rangle|^2,
\end{eqnarray}
\end{widetext}
where $r=+(-)$ denotes the right(left)--moving component, and the
cross terms of oppositely moving components cancel out due to
fermionic anti-commutation relations.

The transition amplitudes at $X_L$ is identical to that at $X_R$.
For simplicity, we consider the case at $X_L=0$ only. The overlap
elements of the many-body occupations
$\bra{\{n\}}=\bra{n_1,n_2,\cdots,n_m,\cdots}$ and
$\ket{\{n'\}}=\ket{n_1',n_2',\cdots,n_m',\cdots}$ are
\begin{equation} \label{eq:app_DotT_psi}
|\langle \{n\} |\psi^\dagger_{r}(x=0)| \{n^\prime\}\rangle|^2
=\frac{1}{2\pi\Lambda}\prod_{m=1}^{\infty}|\langle
n_m|\varphi_{m}|n_m^\prime\rangle |^2,
\end{equation}
where $\varphi_{m}=\exp\left[\lambda_m(b_m+b_m^\dagger)\right]$ with
$\lambda_m=-\frac{i}{\sqrt{gmM}}$ is the bosonized field operator at
an edge of the wire with open boundary conditions (see for instance
Ref. \onlinecite{Mattsson97a}).  Here $g$ is the interaction
parameter, $m$ is the mode index (and the integer momentum of it),
and $M$ is the number of transport sectors; if $M>1$, the
contributions of the different sectors must be multiplied. The
operators $b_m$ and $b_m^\dagger$ denote plasmon annihilation and
creation and $\Lambda$ is a high energy cut-off.

Using the Baker-Haussdorf formula 
\begin{equation}\label{eq:app_DotT_varphi}
\varphi_{m}=\exp\left[\lambda_m(a_m+a_m^\dagger)\right]
=e^{-\frac{\lambda^2}{2}}e^{\lambda_m a_m}e^{\lambda_m
a^\dagger_m},
\end{equation}
and the harmonic oscillator states
\begin{equation} \label{eq:app_DotT_ladder}
|n\rangle= \frac{(a^{\dagger})^{ n}}{\sqrt{n!}}|0\rangle,~~~
\langle n|=\langle 0| \frac{a^{n}}{\sqrt{n!}},
\end{equation}
one can show that, if $n\leq n'$,
\begin{multline} \label{eq:app_DotT_varphi2}
|\braket{n| \varphi|n'}|^2=
e^{|\lambda|^2}\frac{|\lambda|^{2(n-n')}}{n! n'!}\left(\frac{n!}{
(n-n')!}\right)^2 \\ \times\Phi(n+1,n-n'+1;-|\lambda|^2)^2,
\end{multline}
where $\Phi(x,x';z)$ is a degenerate hypergeometric function
defined by \cite{Gradshteyn80a}
\begin{equation}
\Phi(x,x';z)=\left[ \sum_{\ell=0}^{\infty} \frac{z^\ell}{\ell !}
\frac{(x-1+\ell)!}{(x-1)!} \frac{(x'-1)!}{(x'-1+\ell)!} \right].
\end{equation}
If $n'\leq n$, the indices $n$ and $n'$ are exchanged in Eq.
\eqref{eq:app_DotT_varphi2}. The function $\Phi(x,x';z)$ is a
solution of the equation
$$ z\partial^2_z \Phi + (x'-z)\partial_z \Phi-x \Phi=0. $$
By solving this differential equation with the proper
normalization constant, one obtains
$$ \Phi(n+1,n-n'+1;-|\lambda|^2)=
\frac{n'!(n-n')!}{n!} e^{-|\lambda|^2}L_{n'}^{n-n'}(|\lambda|^2),
$$
where $L_{a}^{b}(y)$ is the Laguerre polynomials
\cite{Gradshteyn80a}. In terms of the Laguerre polynomials,
therefore, the transition amplitude \eqref{eq:app_DotT_varphi2} is
written by
\begin{multline} \label{eq:app_DotT_varphi3}
|\langle n_m|\varphi_{m}|n_m^\prime\rangle |^2 =
\frac{e^{-1/gmM}}{(gmM)^{|n_m^\prime-n_m|}}  \frac{n_m^{(<)}
!}{n_m^{(>)} !} \\ \times\left[
L^{|n_m^\prime-n_m|}_{n_m^{(<)}}\left(\frac{1}{gmM}\right)\right]^2
\end{multline}
where $n^{(<)}=\min(n^\prime,n)$ and $n^{(>)} = \max(n^\prime,n)$.

We introduce a high frequency cut-off $m_c\sim k_F L_D/\pi$ to
cure the vanishing contribution due to $e^{-1/gmM}$,
\begin{equation} \label{eq:app_DotT_cutoff}
\frac{1}{2\pi\Lambda} \prod_{m=1}^{m_c} e^{-1/4mg} \rightarrow
\frac{1}{2L_D}\left(\frac{\pi \Lambda_c}{L_D}\right)^\alpha,
\end{equation}
where the exponent is $\alpha=(g^{-1}-1)/M$. We arrive at the
desired form of the on-dot transition matrix elements,
\begin{multline} \label{eq:app_DotT_2}
|\{n^\prime\}|\Psi^\dagger_{a}(X_\ell)|\{n\}\rangle|^2
=\frac{1}{L_D}\left(\frac{\pi \Lambda}{L_D}\right)^{\alpha}
\prod_{\nu}  \prod_{m=1}^{\infty}\\ \times \left(\frac{1}{g_\nu
mM}\right)^{|n_m^\prime-n_m|} \frac{n_m^{(<)} !}{n_m^{(>)} !}
\left[ L^{|n_m^\prime-n_m|}_{n_m^{(<)}}\left(\frac{1}{g_\nu
mM}\right)\right]^2
\end{multline}

\section{Universal occupation probability \label{app:P_univ}}

In this appendix, we derive the universal distribution of the
occupation probability Eq. \eqref{eq:P_univ}.

 Since the occupation
probability of the plasmon many-body states is a function of the
state energy in the strong interaction regime, we introduce the
dimensionless energy $n=\sum_m m\cdot n_m$ of the state with
$\{n\}=(n_1,n_2,\cdots,n_m,\cdots)$ plasmon occupations. Excluding
the zero mode energy, therefore, the energy of the state $\{n\}$
is given by $E_D(\{n\})= E_D(n)= n\veps_p$ with the state
degeneracy $D(n)$, i.e. the number of many-body states $\{n\}$
satisfying $n=\sum_m m\cdot n_m$, asymptotically following the
Hardy-Ramanujan formula \cite{HardyG18a}
\begin{equation}\label{eq:app_HR}
D(n) \simeq e^{\pi\sqrt{2n/3}}/(4\sqrt{3}n).
\end{equation}
We denote $n_{sd}$ by the corresponding dimensionless bias voltage
$eV=n_{sd} \veps_p$.

We obtain an analytic approximation to the occupation probability
$P(n)$ at zero temperature by setting the on-dot transition
elements in (\ref{eq:Laguerre}) to unity and considering the
scattering-in and scattering-out processes for a particular
many-body state $\{n\}$.

The total scattering rates at zero temperature are given by a
simple power-law Eq. \eqref{eq:gamma_T0},
\begin{multline} \label{eq:app_univ_Gamma}
\Gamma(n\tol m)=
\Theta(-n+m-{n_{sd}}/2)(-n+m-{n_{sd}}/2)^\alpha  \\
  +\Theta(-n+m+{n_{sd}}/2)(-n+m+{n_{sd}}/2)^\alpha \\
  \approx \Theta(-n+m-{n_{sd}}/2)(-n+m-{n_{sd}}/2)^\alpha
\end{multline}
where the constant factor in Eq. \eqref{eq:gamma_T0} is set to
unity.

The master equation now reads
\begin{equation}
\frac{\partial}{\partial t} P(n)=\sum_{m} \bigl[ P(m) D(m)\Gamma(m
\rightarrow n) -P(n)D(m)\Gamma(n \rightarrow m) \bigr]
\label{eq:app_univ_master}
\end{equation}
To solve this master equation, we assume an ansatz of a power-law
\begin{equation}\label{eq:app_univ_ansatz}
P(m)=P(n) q_n^{m-n}.
\end{equation}
In the steady-state, master equation \eqref{eq:app_univ_master} in
terms of this ansatz becomes
\begin{multline} \label{eq:app_univ_master2}
\sum_{m= m_i}^{\infty} (m-n+n_{sd}/2)^\alpha D(m) q_n^{m-n}
\\ \approx \sum_{m'=0}^{n+n_{sd}/2}(n-m'+n_{sd}/2)^\alpha D(m'),
\end{multline}
in which the sum in the LHS runs from $m_i=\max(0,n-n_{sd}/2)$,
where $\max(x,x')$ returns larger value of $x$ and $x'$.

Using a saddle point integral approximation
\begin{multline} \label{eq:app_univ_saddle}
\int e^{f(x)} dx \approx e^{f(x_0)}\int
e^{\frac{1}{2}f^{\prime\prime}(x_0)(x-x_0)^2} dx, \\
 \mbox{ if }
f^\prime(x_0)=0,~f^{\prime\prime}(x_0) <0,
\end{multline}
we solve Eq. \eqref{eq:app_univ_master2} to obtain an equation for
$\ln q_n$ and find that for a large $n$,
\begin{equation} \label{eq:app_univ_ln_q}
\exp(z)/z\approx \sqrt{n} \exp(F(n)),
\end{equation}
where
$$z=\frac{|\ln(q)|}{C_\alpha},~~ C_\alpha =
\frac{\alpha+1}{n_{sd}/2}, $$ and $F(n)$ is a slowly varying
function of $n$ for $n\gg 1$,
\begin{equation} \label{eq:app_univ_F}
F(n) = \sqrt{\frac{2}{3}}\frac{\pi}{C_\alpha
\sqrt{n}}+\ln\left[\frac{\sqrt{6}}{\pi}C_\alpha\right]
+\frac{n_{sd}}{n}\left(\frac{1}{4}-\frac{1}{\alpha+1} \right)
\end{equation}

We assume an ansatz for the solution of $z$ in Eq.
\eqref{eq:app_univ_ln_q},
\begin{equation} \label{eq:app_univ_ansatz2}
z=(\ln n)/2 +F(n)+\sqrt{(\ln n)/2+F(n)-K} +\eta
\end{equation}
and find a constant $K$ which minimizes the correction term
$\eta$. Putting this ansatz into Eq. \eqref{eq:app_univ_ln_q} and
solving the equation for $\eta$, we find at $K\approx 0.8$, the
correction term $\eta$ is negligibly small ($\eta\approx 0.01$).

Noting $f(n)=\sqrt{(\ln n)/2+F(n)-K}$ is almost linear function in
the regime of our interest ($3\lesssim n\lesssim 15$), we
linearize it around a value of interest $n=n_0$ (for instance,
$n_0=9$),
$$ f(n)\approx f'(n_0)(n-n_0)+f(n_0), $$
and solve Eq. \eqref{eq:app_univ_ln_q} by ansatz
\eqref{eq:app_univ_ansatz2} with above linearized form;
\begin{equation}\label{eq:app_univ_ln_q2}
-\frac{\ln q_n}{C_\alpha} \approx \frac{1}{2}\ln n+ F(n)+f'(n_0)n+
f(n_0)- f'(n_0) n_0.
\end{equation}

Apply $\partial_m \ln[P(m)]\approx \ln[q_n]$ to Eq.
\eqref{eq:app_univ_ln_q2}, and solve the integral equation for
$\ln[P(n)]$,
\begin{multline}\label{eq:app_univ_lnP}
\ln[P(n)] \approx -C_\alpha \int^n d n \Big[(\ln n)/2+F(n) \\
+ f'(n_0) n+f(n_0)-n_0f'(n_0)\Big] .
\end{multline}
The leading order approximation $P^{(0)}(n)$ to the probability
$P_{\mathrm{univ}}(n)$ of the average occupation from this
integral results in Eq. \eqref{eq:P_univ}
\begin{equation}
P^{(0)}(n) =Z^{-1}~ n^{-\frac{3}{2}\frac{\alpha+1}{n_{sd}}n}
=Z^{-1}\cdot e^{-\frac{3}{2}\frac{\alpha+1}{n_{sd}}n\log n},
\label{eq:app_univ_P0}
\end{equation}
where $Z$ is the partition function.

A more accurate approximation $P^{(1)}(n)$ can be derived
by solving the integral Eq. \eqref{eq:app_univ_lnP} to a higher degree
of precision, which yields
\begin{widetext}
\begin{multline} \label{eq:app_univ_P1}
P^{(1)} =Z^{-1}~ \left(\sqrt{n}
\left(n+\frac{n_{sd}}{2}\right)\right)^{-C_\alpha n/2}\times
\left(\frac{\sqrt{6}}{\pi} C_\alpha\right)^{-C_\alpha n}
\exp\left(-C_\alpha n \left[F(n_0)-K+\frac{\ln
3}{2} -\frac{3}{4}\right] \right) \\
\times \exp\left(-\frac{4\pi}{\sqrt{6}}\sqrt{n}
\right)\left(\frac{n-n_{sd}/2}{n+n_{sd}/2}\right)^{-2n/n_{sd}}
\left(n+\frac{n_{sd}}{2}\right)^{-(\alpha+1)/2}
\left(n^2-\left(\frac{n_{sd}}{2}\right)^2\right),
\end{multline}
\end{widetext}
where $n_0=9$ is used and $f(n_0)\approx f(n_0)^2$ with minor
correction is utilized for formal simplicity.  Note the first term
with normalization constant $Z$ approaches $P^{(0)}(n)$ in Eq.
\eqref{eq:app_univ_P0} as $n/n_{sd}\to 0$, noticing
$C_\alpha=2(\alpha+1)/n_{sd}$. The integral equation
\eqref{eq:app_univ_lnP} can be solved even without approximating
on $f(n)$, with the expense of more cumbersome appearance of
$P(n)$.

\end{document}